\documentclass[aps,pre,twocolumn,groupedaddress]{revtex4}
\usepackage{graphics}% Include figure files
\usepackage{graphicx}% Include figure files

\begin{document}

%\preprint{}

\title{A self-consistent model of the plasma staircase and nonlinear Schr\"odinger equation with subquadratic power nonlinearity}


%\affiliation can be followed by \email, \homepage, \thanks as well.
%\author{Alexander~V.~Milovanov${}^{1,2}$, Guilhem~Dif-Pradalier${}^{3}$, and Jens~Juul~Rasmussen${}^{4}$}
%\email[]{Alexander.Milovanov@phys.uit.no}
%\homepage[]{www.phys.uit.no}
%\thanks{}
%\altaffiliation{Also at: Department of Space Plasma Physics, Space Research Institute, Russian Academy of Sciences, Profsoyuznaya 84/32, 117997 Moscow, Russia}

%\email[]{jens.juul.rasmussen@risoe.dk}
%\thanks{}

%\author{Alexander~V.~Milovanov,${}^{1,2,3}$~Guilhem~Dif-Pradalier,${}^{4}$~and~Jens~Juul~Rasmussen${}^{5}$}

%\affiliation{${}^1$ENEA National Laboratory, Centro~Ricerche~Frascati, I-00044 Frascati, Rome, Italy}
%\affiliation{${}^2$Max-Planck-Institut f\"ur Physik komplexer Systeme, D-01187 Dresden, Germany}
%\affiliation{${}^3$Space Research Institute, Russian Academy of Sciences, 117997 Moscow, Russia}
%\affiliation{${}^4$CEA-Cadarache, IRFM, F-13108 Saint-Paul-lez-Durance cedex, France}
%\affiliation{${}^5$Physics Department, Technical University of Denmark, DK-2800 Kgs.~Lyngby, Denmark}

\author{Alexander V. Milovanov}
\affiliation{ENEA National Laboratory, Centro~Ricerche~Frascati, I-00044 Frascati, Rome, Italy}
%\affiliation{Max-Planck-Institut f\"ur Physik komplexer Systeme, 01187 Dresden, Germany}
\affiliation{Space Research Institute, Russian Academy of Sciences, 117997 Moscow, Russia}

\author{Jens Juul Rasmussen}
\affiliation{Physics Department, Technical University of Denmark, DK-2800 Kgs.~Lyngby, Denmark}

\author{Guilhem Dif-Pradalier}
\affiliation{CEA-Cadarache, IRFM, F-13108 Saint-Paul-lez-Durance cedex, France}


%\date{\today}

\begin{abstract} A new basis has been found for the theory of self-organization of transport avalanches and jet zonal flows in L-mode tokamak plasma, the so-called ``plasma staircase" (Dif-Pradalier {\it et al.}, Phys. Rev. E, {\bf 82}, 025401(R) (2010)). The jet zonal flows are considered as a wave packet of coupled nonlinear oscillators characterized by a complex time- and wave-number dependent wave function; in a mean-field approximation this function is argued to obey a discrete nonlinear Schr\"odinger equation with subquadratic power nonlinearity. It is shown that the subquadratic power leads directly to a white L\'evy noise, and to a L\'evy-fractional Fokker-Planck equation for radial transport of test particles (via wave-particle interactions). In a self-consistent description the avalanches, which are driven by the white L\'evy noise, interact with the jet zonal flows, which form a system of semi-permeable barriers to radial transport. We argue that the plasma staircase saturates at a state of marginal stability, in whose vicinity the avalanches undergo an ever-pursuing localization-delocalization transition. At the transition point, the event-size distribution of the avalanches is found to be a power-law $w_\tau (\Delta n) \sim \Delta n^{-\tau}$, with the drop-off exponent $\tau = ({\sqrt{17}} + 1)/{2} \simeq 2.56$. This value is an exact result of the self-consistent model. The edge behavior bears signatures enabling to associate it with the dynamics of a self-organized critical  (SOC) state. At the same time the critical exponents, pertaining to this state, are found to be inconsistent with classic models of avalanche transport based on sand-piles and their generalizations, suggesting that the coupled avalanche-jet zonal flow system operates on different organizing principles. The results obtained have been validated in a numerical simulation of the plasma staircase using flux-driven gyrokinetic code for L-mode Tore-Supra plasma.   
%We resolve an existing question concerning localization-delocalization of plasma avalanches by jet zonal flows in a tokamak using a novel theoretical scheme: that of the nonlinear Schr\"odinger equation with subquadratic power nonlinearity. In the theoretical model the event-size distribution of the avalanches follows a asymptotic power-law $w_\tau (\Delta n) \sim \Delta n^{-\tau}$ for $\Delta n \rightarrow +\infty$, with the exponent $\tau = ({\sqrt{17}} + 1)/{2} \simeq 2.56$. This value is obtained self-consistently based on the idea of self-organization of the staircase into a marginally stable state, using a comblike potential function for the jet flows and the formalism of L\'evy-fractional Fokker-Planck equation. Similarities and differences of the staircase dynamical system with regard to both ordinary self-organized criticality (SOC) systems and the nonlinear Anderson problem are discussed. The results obtained have been validated in a numerical simulation of the staircase dynamics using flux-driven gyrokinetic code for L-mode Tore-Supra plasma.    
\end{abstract}

%\pacs{05.45.Mt, 72.15.Rn, 42.25.Dd, 05.45.-a}
%\keywords{Anderson localization \sep algebraic nonlinearity \sep mean-field percolation}

\maketitle

\section{Introduction} 
Recently, due to the high-resolution, ultrafast sweeping reflectometry schemes employed in the fusion research, %and the demand for smart plasma diagnostics systems, 
there has been increasing attention both theoretically and experimentally on the issues related with the propensity of toroidally confined L-mode plasma to spontaneously generate micro-barriers to radial transport as a result of plasma self-organization. Often such barriers are found to occur in quasiregular patterns of highly concentrated, multiple jet zonal flows interspersed with broader regions of turbulent (typically, avalanching) transport \cite{DF2010,DF2015,DF2017,Horn2017}. The phenomenon$-$illustrated numerically in Fig.~1 with the aid of a flux-driven gyrokinetic code \cite{Sarazin}$-$has come to be known as the plasma staircase and was so named \cite{DF2010} after its celebrated planetary analogue \cite{McIntyre}.

\begin{figure}[t]
\includegraphics[width=0.46\textwidth]{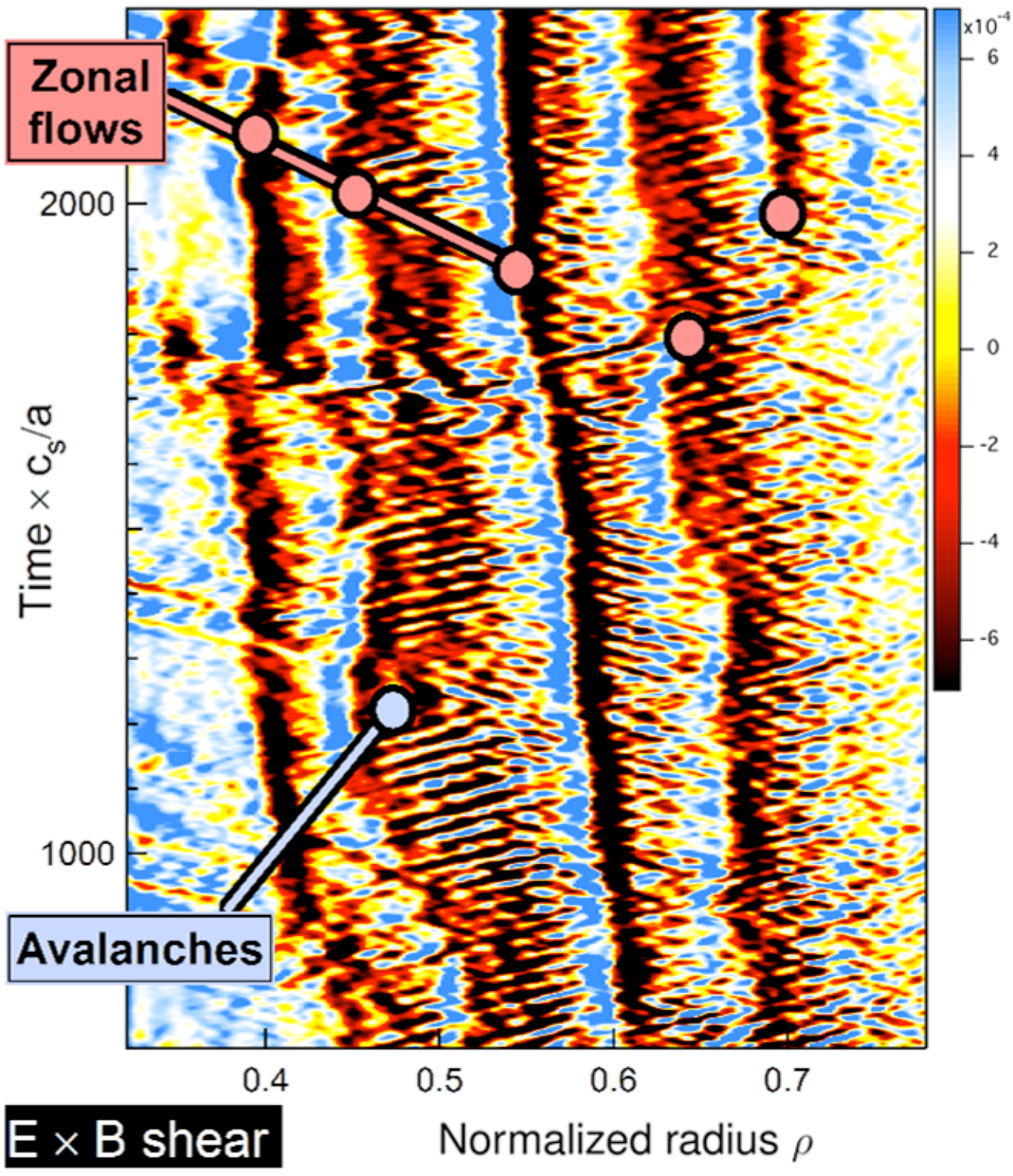}
\caption{\label{} A detail of the plasma staircase in a flux-driven gyrokinetic computation using the \textsc{Gysela} code \cite{Sarazin}. The plasma parameters mimic those of the Tore Supra shot $\#$ 45 511 \cite{DF2015}. The jet zonal flows are thick, nearly vertical bands slightly inclined from bottom-right to upper-left. We have used color to depict the shear of the ${\bf E} \times {\bf B}$ flow: The color scale is expressed in ion cyclotron frequency units, with the color legend summarized on the right-hand side of the figure (blue (light gray) signifies positive values of the shear, and dark/brownish signifies negative values). At the sharp transition between the colors the shear is vanishing. The flat areas between the jets have mixed color. The avalanches are thin, fine-scale structures (brushlike) elongated in radial direction. Concerning the units, ``$\rm a$" is the tokamak minor radius, ``$\rm c_s$" is the ion acoustic speed, and $\rho$ is the dimensionless radial coordinate.}
\end{figure}

The physics of the plasma staircase is of interest from both a fundamental scientific perspective and for the practical realization of fusion energy. From a scientific perspective, the nonlinear dynamics of the plasma staircase occupies an interesting niche where micro- and meso-scale nonlinearities can appear on an equal footing \cite{DF2017,Horn2017}. From a practical perspective, the periodic dynamical patterning due to the plasma staircase offers a unique environment to control the avalanche activity by fine tuning the shape and the radial positions of the barriers \cite{Horn2017,PRE18}. These practical aspects are dictated by the understanding that the avalanche transport may have a deteriorating effect on the confinement properties of thermal plasma and charged fusion products \cite{Zonca06,Heid}, while significant losses could be detrimental. It is therefore a crucial issue to understand the behavior of the coupled staircase-avalanching system and the way the avalanches may be contained within the steps of the barriers.  

Although the plasma staircase is a relatively new topic for fusion, it already enjoys an exciting history behind: The phenomenon was discovered experimentally \cite{DF2015} on the Tore Supra tokamak following its very precise theoretical prediction in Ref. \cite{DF2010}$-$a rare, classic circumstance when the discovery is made {\it au bout de sa plume}, if the celebrated art phrase due to Fran\c{c}ois Arago \cite{Para} is appropriate here. By the time this paper is being written, the natural tendency of L-mode plasma to generate staircase structures proves to be an established fact \cite{DF2017,Horn2017}, it has been confirmed computationally using different numerical codes \cite{Rath,Ghendrich2018,Wang2018,Weikl2018,Qi2019,Ashourvan2019}; discussed theoretically$-$especially invoking flux-gradient time delay, flux landscape bistability or wave trapping  \cite{Kosuga2013,Kosuga2014,Ashourvan2017,Guo2019,Malkov2019,Garbet21}; and observed experimentally other than on Tore Supra also on DIII-D \cite{Ashourvan2019}; KSTAR \cite{Choi2019}; and lately in HL-2A L-mode discharges \cite{HL2A}. 

Our purpose here is to describe a new theoretical framework, whose mathematical foundations have been spelled out in Refs. \cite{Skokos,PRE14,DNC,PRE19}, concerning the behavior of the staircase system near a marginally stable state, where the event-size distribution of the avalanches might be obtained using general arguments, but where nevertheless known approaches based on the assumptions of locality and next-neighborlike interactions do not apply. % self-organized criticality (SOC) \cite{Bak87,Tang,Works} do not apply. 
The key element to our model is the concept of nonlinear Schr\"odinger equation (NLSE) with subquadratic power nonlinearity \cite{PRE14,PRE19}, based on which we could demonstrate the existence of an attracting steady state for the coupled avalanche-jet zonal flow system, and to predict the statistical characteristics of this state. The results of this analysis strongly suggest that the plasma staircase operates as a complex system in a self-organized critical state (SOC) \cite{Bak87,Tang}. %This being said, the statistical distributions of plasma avalanches are found to be not compatible with the usual sand-pilelike models of SOC, indicating a different universality class. 
In general, we could argue that NLSE with subquadratic power nonlinearity offers a fertile basis to study the self-organization phenomena involving SOC, due to the nonlinear twists it carries. 

The main practical result is that the event-size distribution contains an asymptotic ``fat" power-law tail describing the significant likelihood of extreme avalanches. The rate of decay of the power law is found, however, to be inconsistent with familiar models of avalanche transport based on sand-piles and their modifications \cite{Bak87,Tang,Zhang,Kadanoff98},
%self-organized criticality (SOC) \cite{Bak87,Tang,Works} using sand-piles and their generalizations,  %noticeably steeper, though, than what one would expect based on a SOC hypothesis, 
suggesting that the coupled avalanche-zonal flow system works on different organizing principles. Aside from answering a long open question concerning the localization-delocalization of plasma avalanches \cite{DF2017,PRE18}, the understanding of these principles may be of added interest, as it introduces a new approach to study the self-organization of systems with many interacting degrees of freedom.  

The paper is organized as follows. We formulate the NLSE model first (Sec.~II), followed by a demonstration that this model leads directly to the white L\'evy noise, and to a L\`evy-fractional Fokker-Planck equation (FFPE) for particle transport in the radial direction (Sec.~III). Both test particles and a self-consistent transport model are considered. An anxious reader willing to proceed directly to FFPE may skip the derivation of dynamical Eq.~(\ref{4s+}) beginning from Eq.~(\ref{3s}). The steady-state solutions of FFPE are obtained in Sec.~IV for both small and large wave numbers. In Sec.~V we derive the event-size distribution of avalanches in vicinity of the steady state. Also in Sec.~V we validate results using flux-driven gyrokinetics and address the statistical case of extreme avalanches based on the notion of SOC. We have collected our conclusions in Sec.~VI. For the reader's convenience we left to the Appendix a derivation of FFPE for Markov stochastic processes with nonlocal kernel, which is a theory problem {on its own}.

\section{NLSE with subquadratic power nonlinearity} 

We assume that the jet zonal flows are so narrow and concentrated that one might speak about their spatially localized positions in the direction of the tokamak minor radius (see Fig.~1). In order words, the radial widths of the jets are taken to be much thinner than the spacing between them. This condition is generally very well satisfied in the plasma L-mode \cite{DF2017,Horn2017}. With this implication in mind, let us assign to each jet a radial position coordinate, $j$, then consider the jets in the poloidal cross-section, neglecting eventual toroidal drifts (see Fig.~2a,c). In this cross-section, the jets will be represented by closed contours along which the flow is (almost) periodic. The period of the flow is defined as $T_j = l_j / u_{E}$, where $u_{E} = |{\bf E} \times {\bf B}| / B^2$ is the familiar ${\bf E} \times {\bf B}$ velocity, and $l_j$ is the length of the contour. We consider each periodic flow as a nonlinear oscillator, characterized by the nonlinear frequency $\omega_j = 2\pi / T_j$ and the associated wave number $k_j = 2\pi / l_j$. In this interpretation, the staircase is none other than a wave packet of coupled nonlinear oscillators, each with its own identity parameter, $j$. We envisage this wave packet being broad enough in that it contains a large number of the individual jets. The coupling between the jets is provided by their nonlinear interaction, which is mediated by the avalanches. In a self-regulating, nonlinear plasma system, that would be a rather efficient mechanism, since the avalanches, absorbed by the transport barriers, deliver momentum to the poloidal flows (via the turbulent Reynolds stress), which in turn enhances the strength of the barriers \cite{Xu}.   

\subsection{Description of the model}

To characterize, from a most general perspective, the nonlinear dynamics of a wave packet of coupled nonlinear oscillators, the jet zonal flows, one might invoke the analytical scheme of the nonlinear Schr\"odinger equation, or NLSE \cite{Segur,Sulem}, also known as the Gross-Pitaevskii equation \cite{Gross,Leggett}. The time-dependent Gross-Pitaevskii equation describes the dynamics of initially trapped Bose-Einstein condensates and is shown to be an exact equation in the dilute limit \cite{Edros,Lieb}. For many-body bosonic systems, the NLSE is a mean-field approximation where the term proportional to the probability density $|\psi|^2$ represents the interaction between the atoms.

\begin{figure}
\includegraphics[width=0.52\textwidth]{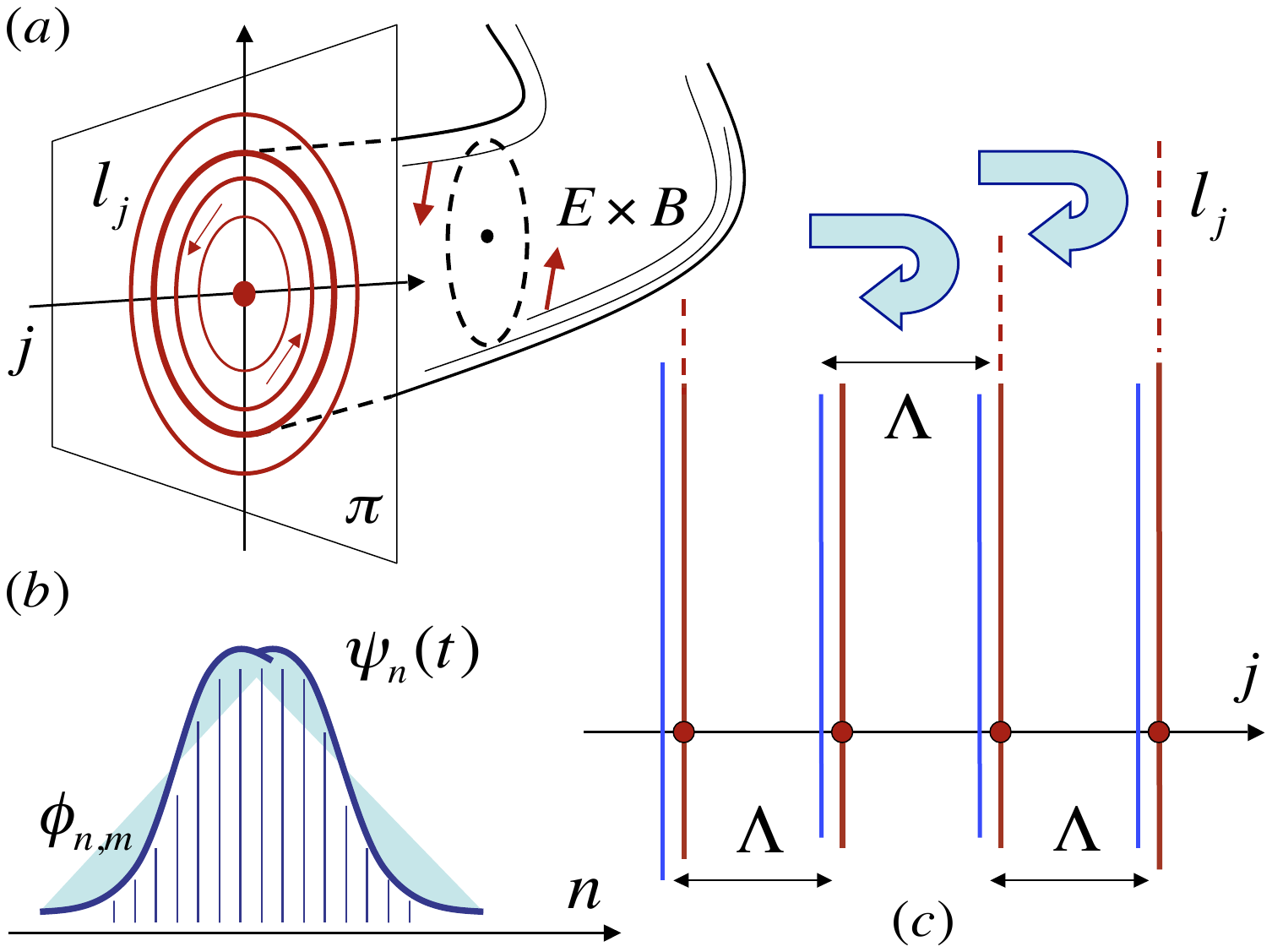}
\caption{\label{} (a) A birds-eye view of the jet zonal flows and their poloidal cross-section (poloidal plane marked with the letter $\pi$). (b) A schematic representation of the staircase wave function, $\psi_n (t) = \psi (n, t)$. Vertical lines mimic the eigenfunctions of the linear problem, $\phi_{n,m}$. (c) The discrete structure of jet zonal flows at the short view and the definition of the position coordinate $j$. The distance between the jets (velocity extrema) is $\Lambda$ and is estimated as the Rhines length for electrostatic drift-wave turbulence, i.e., $\Lambda \sim\Lambda_{\rm Rh}$ (Sec.~IV\,D). %The Rhines length behaves as a square root of the $\bf E \times \bf B$ drift, that is, $\Lambda_{\rm Rh} \propto \sqrt{u_{E}}$, similarly to its fluid analog \cite{McIntyre}. 
The radial position is tracked by the variable $n$, which also characterizes the energy spectrum of the staircase dynamical system under the NLSE approximation. %While we tacitly associate $n$ with $j$, it should be noted that $n$ has a somewhat different implication, since it directly characterizes the energy spectrum of the staircase system under the NLSE approximation. 
Top: U-turn arrows illustrate the avalanches confined between the staircase steps.
}
\end{figure}

Next, we argue (and confirm through results) that the self-organization phenomena pertaining to the plasma staircase require a modified form of NLSE in which the probability density $|\psi|^{2}$ is replaced by a subquadratic power nonlinearity $|\psi|^{2s}$, where $0 < s < 1$ is a power exponent and tunes the nonlinear interaction mechanism. This modified form of NLSE has been considered in Ref. \cite{PRE19} for the destruction of Anderson localization in quantum nonlinear Schr\"odinger lattices with disorder. 

In the nonlinear Anderson problem the subquadratic nonlinearity arises because the nonlinear interactions among the waves might be subject to a competing nonlocal ordering (such as, for instance, the stripy ordering \cite{Kivel,PRB02}, etc.), leading the constituent linear waves to interfere with themselves \cite{PRE14,PRE19}. This destructive self-interference might be either complete eliminating the dependence on the modulus field (for $s=0$) or partial (for $0 < s < 1$), and is parametrized by the subquadratic power $2s < 2$. No competing ordering (no self-interference) is assumed to take place for the quadratic power nonlinearity, with $s=1$. In magnetically confined fusion plasma, a structural disorder similar to the disorder in the Anderson problem might occur thanks to the presence of a low-frequency, electrostatic micro-turbulence (e.g., Refs. \cite{Wagner,Reuss,JJR,Basu}; Ref. \cite{Horton} for review); in the meantime, the competing nonlocal ordering could be associated with spontaneous occurrence of the jet zonal flows \cite{Itoh} or staircase self-organization \cite{DF2015,DF2017,Horn2017}, suggesting a similar dynamical description. Note that the drift waves are simultaneously a source for the disorder and the driving mechanism for the zonal flows. With these implications in mind, we introduce a discrete NLSE of the form 
%can rely on an incomplete overlap between the constituent waves, such that the $s$ values being close to zero correspond to non-overlapping waves; $s$ being close to 1 correspond to the familiar four-wave overlap in the dilute limit \cite{Sh93,PS,Fishman,Flach}, and the general situation being that of a non-integer parameter $s$ taking values from the interval $0 < s < 1$, with larger $s$ corresponding to a greater degree of the overlap. An incomplete overlap may arise due to e.g., repulsive interaction between the components of the wave field subject to complex self-interference processes \cite{PRE14,PRE19}. We expect the incomplete overlap to be a characteristic of strongly nonlinear regimes in which the dilute limit may be relaxed. More explicitly, we consider a discrete NLSE of the form 
\begin{equation}
i\hbar\frac{\partial\psi_n}{\partial t} = E_n \psi_n + \beta |\psi_n|^{2s} \psi_n + V (\psi_{n+1} + \psi_{n-1}),
\label{1} % Eq.~(\ref{1})
\end{equation}
where $\psi_n = \psi (n, t)$ is a complex wave function and describes the plasma staircase as a compound system of coupled nonlinear oscillators; $n$ is the discrete coordinate and is associated with the radial direction in a tokamak; $\beta$ characterizes the strength of nonlinearity (below for definiteness $\beta > 0$); $0 < s < 1$ absorbs the effect of competing ordering on wave-wave interactions; $V$ is transition matrix element; $E_n$ are on-site energies; and the total probability is normalized to unity: $\sum_n |\psi_n|^{2} = 1$. In what follows, $\hbar = 1$ for simplicity; thus, the energy coincides with the frequency. For $\beta \rightarrow 0$, the staircase decays into a set of loosely connected eigenstates, i.e., (almost) noninteracting jet flows, whose eigenfunctions are exponentially localized, the localization length being much smaller than the spacing between the jets. Note that we have introduced $n$ instead of $j$ to be the position coordinate in the NLSE model (see Fig.~2b,c). A reason for that is that $n$ bears a somewhat different implication in that it directly characterizes the energy spectrum of the staircase system under the NLSE approximation. We assume that the spectrum $E_n$ is discrete and dense, for the positions of the jet zonal flows must correspond to rational values of the tokamak safety factor. Note that the safety factor \cite{Safety} is usually a function of radius, implying that the energy spectrum $E_n$ might be actually very broad, consistently with the above assumptions. 

The background theory for NLSE~(\ref{1}) refers to wave processes with competition between dispersion, randomness, and nonlinearity (e.g., Refs. \cite{Jens95,Christ,Gaid,UFN,Clement,Shapiro,Zonca15}). A large body of work promoting a reduced equation with quadratic power nonlinearity ($s=1$) is documented in Refs. \cite{Sh93,PS,WangSt,Fishman,Flach,Skokos09,Iomin,EPL,PRE17,Iomin2019}. A generalization to subquadratic powers, with $s < 1$, was formulated in Refs. \cite{PRE14,DNC,PRE19}. Superquadratic nonlinearities have been considered in Refs. \cite{Skokos,DNC,PRE14}. Formally, the model in Eq.~(\ref{1}) coincides with the NLSE model introduced in Ref. \cite{PRE19}, but with that simplifying element that we do not assume that the field $\psi_n$ is quantum. That is, our $\psi_n = \psi (n, t)$ is just a complex wave function, {not} an operator wave function as in Ref. \cite{PRE19} (as well as in Refs. \cite{Q-Iom,PRE17}, where the quantum case $s=1$ was analyzed). The general properties of a class of nonlinear Schr\"odinger equations with both sub- and superquadratic nonlinearities were reviewed in Ref. \cite{Scr33}, where one also finds conditions for existence, uniqueness, and stability of solitary wave solutions, along with conditions for blow-up and global existence for the Cauchy problem.  

\subsection{Expanding the staircase wave function over the eigenfunctions of the linear problem}

Focusing on the nonlinear dynamics of the plasma staircase, because the jet zonal flows are strongly localized around their radial positions, $j$, it is convenient to expand the wave function $\psi_n$ over the eigenfunctions of the corresponding linear problem that would result if the oscillators in NLSE~(\ref{1}) were decoupled from each other. The Hamiltonian of the linear problem is obtained by letting $\beta\rightarrow 0$ in Eq.~(\ref{1}), leading to 
\begin{equation}
\hat H_{L} \psi_n = E_n \psi_n + V (\psi_{n+1} + \psi_{n-1}), 
\label{1-LL} % Eq.~(\ref{1-LL})
\end{equation} 
which is easily seen to be the familiar Anderson Hamiltonian in the tight-binding approximation \cite{And} (except that we do not assume that the energies $E_n$ are necessarily random). The eigenfunctions of the linear problem, $\phi_{n,k}$, are defined by $\hat H_{L} \phi_{n,k} = \omega_k \phi_{n,k}$, where $\omega_k$ are the respective eigenfrequencies, and $k = 0,\pm 1,\pm 2,\dots$ is an integer counter. We argue that the functions $\phi_{n,k}$ form a full basis of orthogonal eigenfunctions and as such might be chosen as the basis functions. The orthogonality of $\phi_{n,k}$ is a direct consequence of strong spatial localization of the staircase jets, on the one hand, and of the discrete character of NLSE~(\ref{1}), on the other hand. Indeed it is found in direct numerical simulations of the discrete NLSE with arbitrary power nonlinearity \cite{Blow-93} that the discrete equation exhibits localization in regimes where blow-up cannot occur in the continuum system. From general studies it follows that the blow-up occurs for $s \geq 2$, whereas the regimes with $s < 2$ are unconditionally stable supporting localization \cite{Blow-93,Scr33}. The latter include the subquadratic power case $s < 1$, of main interest here. Without loss in generality, we may consider that the eigenfunctions $\phi_{n,m}$ are normalized to unity, then them being mutually orthogonal would imply 
\begin{equation}
\sum _n \phi^*_{n,m}\phi_{n,k} = \delta_{m,k},
\label{Kro} % Eq.~(\ref{Kro})
\end{equation}
where $\delta_{m,k}$ is Kronecker's delta, and the star denotes complex conjugate. Complementing the discrete character of NLSE~(\ref{1}) is again the argument that the jet zonal flows are situated at the rational values of the tokamak safety factor. Because rational numbers are everywhere dense in real numbers, the allowed radial positions of the staircase jets are everywhere dense in the position coordinate $j$ (and hence are everywhere dense in the tokamak safety factor, which is a function of $j$). Then being dense in the safety factor they would correspond to a set of functions $\phi_{n,k}$ that form a {\it full} basis of orthogonal eigenstates (similarly to the Anderson problem for which the full basis is known to exist and is well-defined \cite{Pastur}). To this end, using the functions $\phi_{n,k}$ as basis functions, we might now expand 
\begin{equation}
\psi_n = \sum_m \sigma_m (t) \phi_{n,m},
\label{2} % Eq.~(\ref{2})
\end{equation}
where $\sigma_m (t)$ are time-dependent complex amplitudes, and $m = 0, \pm 1, \pm 2,\dots$ is an integer counter.

\subsection{Dynamical equations for the complex amplitudes $\sigma_m (t)$}

We now obtain a set of dynamical equations for $\sigma_m (t)$. If $s=1$, then the task is relatively straightforward. One needs to substitute Eq.~(\ref{2}) into NLSE~(\ref{1}), multiply both sides by $\phi^*_{n,k}$, and then sum over $n$, using the orthonormality condition in Eq.~(\ref{Kro}). The result is   
\begin{equation}
i\dot{\sigma}_k - \omega_k \sigma_k = \beta \sum_{m_1, m_2, m_3} V_{k, m_1, m_2, m_3} \sigma_{m_1} \sigma^*_{m_2} \sigma_{m_3},
\label{4} % Eq.~(\ref{4})
\end{equation}
where $\omega_k$ are the eigenfrequencies of the linear problem; the coefficients 
\begin{equation}
V_{k, m_1, m_2, m_3} = \sum_{n} \phi^*_{n,k}\phi_{n,m_1}\phi^*_{n,m_2}\phi_{n,m_3}
\label{5} % Eq.~(\ref{5})
\end{equation}
characterize the overlap structure of the nonlinear field; and we have used a dot to denote time differentiation. 

If $s < 1$, then likewise a simple procedure does not exist (attempting to rise a series expansion resulting from $\psi_n\psi_n^*$ into a fractional power leads one to fight with a chimera). Even so, here we might use a different tack already nailed down in Ref. \cite{PRE19}, using the formalism of Diophantine equations and the notion of the backbone map, introduced in Ref. \cite{PRE14}. The procedure is as follows. 

First of all, we need to discuss what is to be meant by the power $2s$ of the modulus function $|\psi_n|$, and we define this power as the power $s$ of the probability density $|\psi_n|^2$, i.e., $|\psi_n|^{2s} \equiv (|\psi_n|^2)^{s}$. %(consistently with the fact that the modulus function is but the square root of the density). 
Using Eq.~(\ref{2}), we have  
\begin{equation}
|\psi_n|^{2s} = (\psi_n\psi_n^{*})^s = \left[\sum_{m_1,m_2} \sigma_{m_1} \sigma^{*}_{m_2} \phi_{n,m_1}\phi^*_{n,m_2}\right]^s.
\label{2s} % Eq.~(\ref{2s})
\end{equation}
Mathematically, it is convenient to consider the power nonlinearity on the right-hand side of Eq.~(\ref{2s}) as a functional map
\begin{equation}
\hat \mathrm{F}_s:\{\phi_{n,m}\}\rightarrow \left[\sum_{m_1,m_2} \sigma_{m_1} \sigma^*_{m_2} \phi_{n,m_1}\phi^*_{n,m_2}\right]^s
\label{3s} % Eq.~(\ref{3s})
\end{equation}
from the complex vector field $\{\phi_{n,m}\}$ into the scalar field $(|\psi_n|^{2})^s$. It is noticed that the map in Eq.~(\ref{3s}) is positive definite, and that it contains a self-affine character in it, such that by stretching the basis vectors by a stretch factor $\lambda$ the value of $\hat \mathrm{F}_s$ is renormalized (multiplied by $|\lambda|^{2s}$). We have, accordingly,  
\begin{equation}
\hat \mathrm{F}_s\{\lambda \phi_{n,m}\} = |\lambda|^{2s} \hat \mathrm{F}_s\{\phi_{n,m}\}.
\label{4s} % Eq.~(\ref{4s})
\end{equation}

\subsubsection{The multinomial expansion and Diophantine equations in the leading order}

For any nonnegative integer $s$, the power nonlinearity in Eq.~(\ref{2s}) can be expanded in a multinomial series \cite{Stegun} to give 
\begin{equation}
|\psi_n|^{2s} = \sum_{\sum {q_{m_1,m_2}} = s}\mathcal{C}_s^{\dots q_{m_1,m_2}}\prod_{m_1, m_2} [\xi_{m_1,m_2}]^{q_{m_1,m_2}},
\label{Multinom} % Eq.~(\ref{Multinom})
\end{equation}
where 
\begin{equation}
\mathcal{C}_s^{\dots q_{m_1,m_2}} = \frac{s!}{\prod_{m_1, m_2} [q_{m_1,m_2}!]}
\label{Coeff} % Eq.~(\ref{Coeff})
\end{equation}
is a multinomial coefficient, the sign $!$ indicates the factorial operation, and we have denoted 
\begin{equation}
\xi_{m_1,m_2} = \sigma_{m_1} \sigma^*_{m_2} \phi_{n,m_1}\phi^*_{n,m_2}
\label{Simp} % Eq.~(\ref{Simp})
\end{equation}
for simplicity. The sum in Eq.~(\ref{Multinom}) is taken over all combinations of nonnegative integer exponents $q_{m_1,m_2}$ such that the sum of all $q_{m_1,m_2}$ is $s$, i.e.,  
\begin{equation}
\sum_{m_1,m_2}{q_{m_1,m_2}} = s.
\label{Sums} % Eq.~(\ref{Sums})
\end{equation}
An analytic continuation of Eqs.~(\ref{Multinom}) and~(\ref{Coeff}) to noninteger values of $s$ can be obtained by extending the factorial function to the gamma function using $m! = \Gamma (m+1)$ and simultaneously relaxing the condition that the exponents in Eq.~(\ref{Sums}) are integer. The latter generalization might be achieved {\it iteratively} starting from a situation according to which there is only one such exponent to be accounted for, then gradually increasing the number of the fractional-valued exponents in Eq.~(\ref{Sums}), aiming to assess their overall effect on the final expansion. 

So in the first iteration Eq.~(\ref{Sums}) can only be satisfied if {\it the} fractional exponent that we are looking at (which is the {\it only} fractional exponent in this case) is equal to $s$ exactly (because the sum of the remaining integer-valued exponents cannot add up to a fractional value). Then Eq.~(\ref{Sums}) demands that the sum of the remaining (integer-valued) exponents is zero, and this is an exact result. Assume it is the exponent $q_{i,j}$ which takes the fractional value, i.e., $q_{m_1,m_2} = s$ for some $m_1 = i$ and $m_2 = j$. Then from Eq.~(\ref{Sums}) one infers
\begin{equation}
\sum_{m_1\ne i,m_2\ne j}{q_{m_1,m_2}} = 0.
\label{Dioph} % Eq.~(\ref{Dioph})
\end{equation}
Equation~(\ref{Dioph}) is a Diophantine equation, which is a polynomial equation for which only integer solutions are sought. Because the exponents $q_{m_1,m_2}$ cannot take negative values, the only way Eq.~(\ref{Dioph}) can be satisfied is by setting {\it all} the exponents $q_{m_1,m_2}$ to zero ($m_1 \ne i$, $m_2\ne j$; the exponent for which $m_1 =i$ and $m_2 = j$ is equal to $s$, i.e., $q_{i,j}=s$). It is understood that the polynomial form in Eq.~(\ref{Multinom}) is {\it homogeneous} in that the sum of the exponents in each term is always $s$, as Eq.~(\ref{Sums}) shows. On the other hand, the property of the homogeneity implies that any term of the polynomial in Eq.~(\ref{Multinom}) is in some sense representative of the whole. That means that there is no particular reason for which to prefer the very specific setting $m_1 = i, m_2 = j$ against other settings when choosing the fractional-valued exponent, $q_{m_1,m_2}$. The net result is that the condition $q_{i,j} = s$ could be satisfied in a countable number of ways within the range of variation of the parameters $m_1$ and $m_2$. Clearly, all such combinations would equally contribute to the series expansion in Eq.~(\ref{Multinom}). Then to account for these contributions one has to sum over the indexes $m_1$ and $m_2$. Eventually, in the first iteration, Eq.~(\ref{Multinom}) is simplified to 
\begin{equation}
|\psi_n|^{2s} = \sum_{{m_1,m_2}} [\xi_{m_1,m_2}]^{s},
\label{Homo} % Eq.~(\ref{Homo})
\end{equation}
where we have considered that 
\begin{equation}
\mathcal{C}_s^{\dots q_{i,j}} = \frac{\Gamma (s+1)}{\Gamma (q_{i,j} + 1)} = \frac{\Gamma (s+1)}{\Gamma (s + 1)} = 1.
\label{First} % Eq.~(\ref{First})
\end{equation}
Substituting $\xi_{m_1,m_2}$ with the aid of Eq.~(\ref{Simp}), from Eq.~(\ref{Homo}) one arrives at 
\begin{equation}
|\psi_n|^{2s} = \sum_{{m_1,m_2}} \sigma^{s}_{m_1} \sigma_{m_2}^{*s} \phi^{s}_{n,m_1}\phi^{*s}_{n,m_2}.
\label{HF} % Eq.~(\ref{HF})
\end{equation}

\subsubsection{Diophantine equations in the second and higher orders}

Turning to the second iteration, we assume that Eq.~(\ref{Sums}) could be satisfied in such a way that the fractional exponents are just two (and only two), while any other exponents are given by the integer numbers. All these exponents must, moreover, be nonnegative to ensure good behavior in the infrared limit when the interaction amplitudes vanish. Because $s < 1$, the only possibility is that the sum of the fractional exponents is $s$, while the sum of the integer-valued exponents is zero. Denoting the fractional exponents as $q_{i_1,j_1}$ and $q_{i_2,j_2}$, one finds that Eq.~(\ref{Sums}) is split into two separate equations, that is, $q_{i_1,j_1} + q_{i_2,j_2} = s$ and the Diophantine equation 
\begin{equation}
\sum_{m_1\ne i_1, i_2}\sum_{m_2\ne j_1, j_2}{q_{m_1,m_2}} = 0,
\label{Dioph2} % Eq.~(\ref{Dioph2})
\end{equation}
from which it is deduced that all integer-valued exponents are equal to zero, i.e., $q_{m_1,m_2} = 0$ for $m_1\ne i_1, i_2$ and $m_2\ne j_1, j_2$. That means that the only non-vanishing exponents that would meaningfully contribute to the product $\prod_{m_1, m_2}$ on the right-hand side of Eq.~(\ref{Multinom}) are those which take the fractional values, i.e., the exponents $q_{i_1,j_1}$ and $q_{i_2,j_2}$. It is understood that the corresponding wave process in Eq.~(\ref{Multinom}) requires nonlinear coupling among 4 waves, in contrast to only 2 waves in the first iteration, and is represented by a term proportional to $[\xi_{i_1,j_1}]^{q_{i_1,j_1}} [\xi_{i_2,j_2}]^{q_{i_2,j_2}}$. If we adopt, for the reasons of formal ordering, that the coupling probability between two waves in Eq.~(\ref{Multinom}) is characterized by a small parameter $\epsilon \ll 1$, then the two-wave process in the first iteration has the order $\mathcal{O} (\epsilon)$, and the 4-wave process in the second iteration has the order $\mathcal{O} (\epsilon^2)$. More generally, in the $l$-th iteration, a nonlinear coupling among as many as $2l\geq 2$ waves is required, leading to a nonlinear process of order $\mathcal{O} (\epsilon^{l})$. The higher the order is, the less probable the corresponding nonlinear process would be (thanks to the exponential decay of the factor $\epsilon^{l}$ with increasing $l$). Confident on this exponentially fast decay, we might arguably propose that the NLSE dynamics in Eq.~(\ref{1}) are governed by the coupling processes in the first order over $\epsilon$$-$mathematically corresponding to a situation when the fractional exponent in Eq.~(\ref{Sums}) is just one and only one. 

The net result is that the original NLSE model in Eq.~(\ref{1}) could be simplified, and a reduced model based on Eq.~(\ref{HF}) might instead be advanced without losing the essential physics of nonlinear interaction. We note in passing that the reduced model in Eq.~(\ref{HF}) is consistent with the idea that NLSE~(\ref{1}) is by itself an approximation, according to which the nonlinearity $|\psi_n|^{2s}$ occurs as a consequence of the coupling process $|\psi_n|^{2s} = |\psi_n|^{s}\times |\psi_n|^{s}$ in the first order. For $s=1$, this approximation is actually quite known in physics \cite{Fishman,Sulem,UFN}.  

\subsubsection{The backbone map}

Similarly to Eq.~(\ref{3s}) above, the model nonlinearity in Eq.~(\ref{HF}) can be considered as a homogeneous map  
\begin{equation}
\hat\mathrm{F}^\prime_s:\{\phi_{n,m}\}\rightarrow \sum_{m_1,m_2} \sigma^{s}_{m_1} \sigma_{m_2}^{*s} \phi^{s}_{n,m_1}\phi^{*s}_{n,m_2}
\label{3ss} % Eq.~(\ref{3ss})
\end{equation}
from the complex vector field $\{\phi_{n,m}\}$ into a scalar field in Eq.~(\ref{3ss}). The map in Eq.~(\ref{3ss}) was introduced for classical waves in Refs. \cite{PRE14,DNC} and later generalized to quantum waves in Ref. \cite{PRE19}. Through these studies it had received a special name, the ``backbone" map, owing to the very specific reductions it offered in the graph space. It was argued that the backbone map preserved (despite the simplifications it carried) if only the sought dynamical properties of the original NLSE model~(\ref{1}) as well as the algebraic structure of $\hat\mathrm{F}_s$ in the sense of Eq.~(\ref{4s}). Note that the maps $\hat\mathrm{F}_s$ and $\hat\mathrm{F}^\prime_s$ are both {\it self-affine}, with the same renormalization property already included in Eq.~(\ref{4s}). That means that the backbone-reduced dynamical model in Eq.~(\ref{HF}) is characterized by the {\it same} scaling behavior of fluctuating observable quantities, and would lead to the {\it same} scaling laws for transport, as the original model in Eq.~(\ref{1}).    

In view of the above, our further analysis will be based on the backbone-reduced NLSE, which is obtained by replacing the original functional map $\hat\mathrm{F}_s$ by the backbone map $\mathrm{F}^\prime_s$ for $s < 1$. Note that $\hat\mathrm{F}_s$ coincides with its backbone in the limit $s\rightarrow 1$. This property illustrates the particularity of the quadratic power case {versus} arbitrary power nonlinearity and has been already discussed in Ref. \cite{PRE14}. 

%\subsection{Backbone-reduced dynamical model}

Multiplying both sides of the backbone-reduced NLSE by $\phi^*_{n,k}$ and summing over $n$ with the aid of the orthonormality condition in Eq.~(\ref{Kro}), after simple algebra one obtains, similarly to Eq.~(\ref{4}), the following dynamical equations for the complex amplitudes $\sigma_{k} (t)$:    
\begin{equation}
i\dot{\sigma}_k - \omega_k \sigma_k = \beta \sum_{m_1, m_2, m_3} V_{k, m_1, m_2, m_3} \sigma^{s}_{m_1} \sigma_{m_2}^{* s} \sigma_{m_3},
\label{4s+} % Eq.~(\ref{4s+})
\end{equation}
where the new coefficients $V_{k, m_1, m_2, m_3}$ are given by (cf. Eq.~(\ref{5}))
\begin{equation}
V_{k, m_1, m_2, m_3} = \sum_{n} \phi^*_{n,k}\phi^{s}_{n,m_1}\phi_{n,m_2}^{* s}\phi_{n,m_3}.
\label{5s+} % Eq.~(\ref{5s+})
\end{equation}
Equations~(\ref{4s+}) correspond to a system of coupled nonlinear oscillators with the Hamiltonian 
\begin{equation}
\hat H = \hat H_{0} + \hat H_{\rm int}, \ \ \ \hat H_0 = \sum_k \omega_k \sigma^*_k \sigma_k,
\label{6} % Eq.~(\ref{6})
\end{equation}
\begin{equation}
\hat H_{\rm int} = \frac{\beta}{1+s} \sum_{k, m_1, m_2, m_3} V_{k, m_1, m_2, m_3} \sigma^*_k \sigma_{m_1}^s \sigma^{*s}_{m_2} \sigma_{m_3}.
\label{6s+} % Eq.~(\ref{6s+})
\end{equation}
Here, $\hat H_{0}$ is the Hamiltonian of noninteracting harmonic oscillators and $\hat H_{\rm int}$ is the interaction Hamiltonian. Note that $\hat H_{\rm int}$ includes self-ineractions through the diagonal elements $V_{k, k, k, k}$. Each nonlinear oscillator with the Hamiltonian   
\begin{equation}
\hat h_{k} = \omega_k \sigma^*_k \sigma_k + \frac{\beta}{1+s} V_{k, k, k, k} \sigma^*_k \sigma_{k}^s \sigma^{*s}_{k} \sigma_{k}
\label{6+h+s} % Eq.~(\ref{6+h+s})
\end{equation}
and the equation of motion 
\begin{equation}
i\dot{\sigma}_k - \omega_k \sigma_k - \beta V_{k, k, k, k} \sigma^s_{k} \sigma^{*s}_{k} \sigma_{k} = 0
\label{eq+s} % Eq.~(\ref{eq+s})
\end{equation}
represents one nonlinear eigenstate in the system: The eigenstates are identified by their wave number $k$, unperturbed frequency $\omega_k$, the degree of self-interference $s$, and the nonlinear frequency shift $\Delta \omega_{k} = \beta V_{k, k, k, k} \sigma^s_{k} \sigma^{*s}_{k}$. Nondiagonal elements $V_{k, m_1, m_2, m_3}$ characterize couplings between each four eigenstates with wave numbers $k$, $m_1$, $m_2$, and $m_3$. Note that the amplitudes raised to the fractional power $0 < s < 1$ are those resulting from the self-interference processes in the presence of competing ordering. Setting $s\rightarrow 1$, from Eqs.~(\ref{4s+}) and~(\ref{5s+}) one recovers the dynamical model in the quadratic power case, i.e., nonlinear Eq.~(\ref{4}) with the coefficients in Eq.~(\ref{5}). 

It is understood that the excitation of a new eigenstate is none other than the spreading of the wave field in wave number space. Resonances occur between the eigenfrequencies $\omega_k$ and the frequencies posed by the nonlinear interaction terms. From the interaction Hamiltonian in Eq.~(\ref{6s+}), the resonance condition is (Ref. \cite{Comm-Res})
\begin{equation}
\omega_k = s \omega_{m_1} - s \omega_{m_2} + \omega_{m_3}.
\label{Res} % Eq.~(\ref{Res})
\end{equation}
For $s = 0$ (linear model), the resonance condition in Eq.~(\ref{Res}) reduces to $\omega_k = \omega_{m_3}$ and is trivial, while for $s = 1$ (quadratic NLSE) it leads directly to Eq.~(10) of Ref. \cite{EPL}, yielding $\omega_k = \omega_{m_1} - \omega_{m_2} + \omega_{m_3}$. When the resonances happen to overlap, the phase trajectories start to switch from one resonance to another on essentially a random basis. As Chirikov \cite{Chirikov} realized, any overlap of resonances would introduce a random ingredient to dynamics together with some transport in phase space. Applying this argument to NLSE~(\ref{1}), one might expect that a transition to chaos would cause the nonlinear field to spread along the domains of chaotic motion. The analysis of Refs. \cite{PRE14,DNC,PRE19} shows, however, that an unlimited spreading of the classical field is only possible for the quadratic power nonlinearity, with $s = 1$, for which the topology of the overlap is such that there might exist a connected escape path to infinity that lies everywhere within the chaotic domain; while for $s < 1$ an escape to infinity would be interrupted by multiple transitions to regular dynamics, so the analogue escape path is disconnected (a purist might think of it as a Cantor set of the Hausdorff dimension $s$)$-$as a result, an unlimited spreading does not really occur unless quantum tunneling effects are included permitting to overcome the discontinuities. We shall return to this point later on in Sec.~V\,F, where the possibility that the staircase transport barriers might diffuse in radial direction in a tokamak is addressed.

\section{L\'evy-fractional Fokker-Planck equation}

The couplings between the oscillators in Eq.~(\ref{4s+}) lead directly to the white L\'evy noise, and to a superdiffusive radial transport of plasma particles with plasma avalanches. The demonstration involves a few steps, the first one is to ascertain that the nonlinear interactions, included in Eq.~(\ref{4s+}), result in a chaotic behavior of the oscillators. This is relatively straightforward, since all that we need to demonstrate at this point is that the so-called stochasticity parameter \cite{ZaslavskyUFN,Sagdeev,Report} is large. We proceed as follows. 

\subsection{The stochasticity parameter $K$}

If the nonlinear field is spread over $\Delta n \gg 1$ states, then the conservation of the total probability 
\begin{equation}
\sum_n \psi_n^{*}\psi_n \simeq \int |\psi_n|^2 d\Delta n = 1
\label{TP} % Eq.~(\ref{TP})
\end{equation}
would imply $|\psi_n|^2 \simeq 1/\Delta n$. In NLSE system~(\ref{1}) the nonlinear frequency shift behaves with the probability density as $\Delta\omega_{\rm NL} = \beta (|\psi_n|^{2})^s$ and for $\Delta n \gg 1$ will scale with the number of states in accordance with $\Delta\omega_{\rm NL} \simeq \beta / (\Delta n)^s$. On the other hand, the distance between the states, $\delta\omega$, goes to zero as $\propto 1/\Delta n$, provided just that the wave-number space is homogeneous. The stochasticity parameter $K \simeq \Delta \omega_{\rm NL} / \delta\omega$ (often referred to as the Chirikov overlap parameter \cite{189}) shows by how much the nonlinear frequency shift is greater than the distance between the resonances (large $K$ values imply that the resonances strongly overlap, leading to chaotic dynamics \cite{Chirikov,ZaslavskyUFN}). Note that the areas of strong resonance overlap might, in general, be highly structured and strongly shaped, that is, while the condition $K \gg 1$ characterizes the density of the overlap, it says nothing about the way the overlapping resonances are folded in the embedding space. As a result, the chaotic motions might occupy only a fraction of the available phase space, with complex internal organization that might be fractal \cite{Sagdeev,Report}. Using $\Delta\omega_{\rm NL} \simeq \beta / (\Delta n)^s$ and $\delta\omega\simeq 1/\Delta n$, one gets 
\begin{equation}
K \simeq \beta (\Delta n)^{1-s} \gg 1.
\label{K-value} % Eq.~(\ref{K-value})
\end{equation}
One sees that the $K$ value generally depends on the number of states (except for the quadratic power case $s=1$). If $s < 1$, then the condition $K \gg 1$ is {\it always} satisfied, provided just that the number of states $\Delta n$ is large enough, i.e., $(\Delta n)^{1-s} \gg 1/\beta$. If $s = 1$, then $K \gg 1$ demands that the nonlinearity parameter $\beta$ be large by itself (independently of $\Delta n$). To this end, if one requires that the wave packet in NLSE~(\ref{1}) is so broad that the condition $\beta (\Delta n)^{1-s} \gg 1$ is satisfied at the time $t = 0$, then one might argue that the consequent NLSE dynamics will be chaotic for {\it all} $t > 0$. Better, the condition $K \gg 1$ will, for $s < 1$, be even improved through dynamics (because the entropy growth would imply that the number of states is a nondecreasing function of $t$). The net result is that the chaotic property is self-improving for $s < 1$ and is self-preserving ($K$ does not depend on $t$) for $s = 1$. So this property will be naturally there either thanks to the initial condition $\beta (\Delta n)^{1-s} \gg 1$ for $t=0$ and $s < 1$, or following the setting $\beta \gg 1$ for $s=1$.  

\subsection{Self-affinity of the backbone}

The next step would be to take a careful look of the functional map in Eq.~(\ref{3s}) (and of its backbone-reduced counterpart in Eq.~(\ref{3ss})), keeping in mind the self-affinity property in Eq.~(\ref{4s}). Indeed the scaling in Eq.~(\ref{4s}) shows that the interactions are self-similar and can be characterized by a fractal measure $d_f = 2s < 2$ (for the moment being we leave it up to the reader how to interpret the wording ``fractal measure"). This, together with the fact that the behavior is chaotic (because the Chirikov parameter, as we have already seen, is large thanks to Eq.~(\ref{K-value})), would tell us that the couplings between the oscillators in Eq.~(\ref{4s+}) generate a stochastic process with fractal organization, posed by the nonlinear interactions. Assume the nonlinear dynamics are such that there is a nonequilibrium steady state attracting phase-space trajectories, and that the oscillators in Eq.~(\ref{4s+}) have evolved into a vicinity of this state. The fact that such a state might at all exist is highly nontrivial, however, we can hypothesize its existence anyway, and try to put in perspective the consequences. One such consequence is that the coupling process must be {\it stationary}; at least, in the absence of large deviations. Then by being both stochastic and stationary this process could be categorized as a random ``noise" process in the limit $t\rightarrow+\infty$ \cite{Kampen} (as an alternative to a ``motion" process). This noise process must, moreover, be stable \cite{Georges} (because it is generated in vicinity of an attracting stable state). But then there is only one such process, which is simultaneously (i) stable, (ii) stationary, (iii) stochastic, and (iv) leads to self-similar fluctuations: the white L\'evy noise. By white L\'evy noise we mean a stationary random process whose time integral is a symmetric $\mu$-stable L\'evy-flight process of index $\mu = 2s$ \cite{Klafter,Ch2007}. Note that the L\'evy flight trajectory can be assigned a Hausdorff (fractal) measure $d_f = \mu = 2s$ \cite{Georges,Klafter}. While the existence of an attracting stable state in the system~(\ref{4s+}) will be a matter of the forthcoming discussion (see Sec.~V\,E), here we nail down our second point, that is, that the nonlinear couplings between the oscillators in Eq.~(\ref{4s+}) are such that they naturally generate a white L\'evy stable noise in vicinity of their attractor. Note that the power $s$ being smaller than 1 matters in this regard, since it produces a whole family of stochastic noise processes parametrized by the index $\mu = 2s$.     

\subsection{The white noise acting on plasma particles}

Our third (and final) point concerns the fact that the white L\'evy noise$-$whose origin, as we argue, is found in the nonlinear interactions between the coupled nonlinear oscillators in Eq.~(\ref{4s+})$-$can naturally act on plasma particles and by doing so can drive superdiffusive transport of these particles in the radial direction in a tokamak. It is understood that NLSE~(\ref{1}) does {\it not} include, as it is, any ``plasma particles" and that the introduction of the dynamical model in Eq.~(\ref{1}) has the only scope of unveiling the mechanism of the noise. This said, we are ready to take the next step and approach the coupled particle-jet oscillator system. The task is twofold and includes two levels of description: test particles and a self-consistent kinetic model. 

\subsubsection{Test particles}

Focusing on the test particles first, we consider a simple system of Langevin equations \cite{Kampen,Georges} for the microscopic motion of the particles under the action of a stochastic (noiselike) force, i.e.,  
\begin{equation}
dn/dt = v;~dv/dt = -\eta v + F_{\mu} (t),
\label{Lvin} % Eq.~(\ref{Lvin})
\end{equation}
where $\eta$ is the plasma viscosity, $n$ is the radial coordinate, $v$ is the particle velocity along $n$, and $F_\mu (t)$ denotes the Langevin source. Based on the above reasoning (Sec.~III\,B), we associate the Langevin source $F_\mu (t)$ with the white L\'evy noise of index $\mu$. As is well known \cite{Klafter,Ch2007,Ch2004,Fog,Jesper,Gonchar,PLA2014}, the Langevin Eq.~(\ref{Lvin}) with the white L\'evy noise $F_\mu (t)$ leads directly to a L\'evy-fractional diffusion equation  
\begin{equation}
\frac{\partial}{\partial t} f (n, t) =  D_\mu \frac{\partial^\mu}{\partial |n|^\mu} f (n, t),
\label{LFDiff} % Eq.~(\ref{LFDiff})
\end{equation} 
where $f = f (n, t)$ is the probability density to find a test particle at point $n$ at time $t$;
\begin{equation}
\frac{\partial^\mu}{\partial |n|^\mu} f (n, t) = \frac{1}{\Gamma_\mu}\frac{\partial^2}{\partial n^2} \int_{-\infty}^{+\infty}\frac{f (n^\prime, t)}{|n-n^\prime|^{\mu - 1}} dn^\prime
\label{Def+} % Eq.~(\ref{Def+})
\end{equation} 
is the Riesz fractional derivative of order $1 < \mu < 2$ (for the various aspects of fractional differentiation and integration see, e.g., Refs. \cite{Samko,Podlubny}; for the Cauchy special case $\mu = 1$ see Refs. \cite{Mainardi,Ch2004,Ch2007}); $\Gamma_\mu$ is the normalization parameter and is given by $\Gamma_\mu = - 2\cos(\pi\mu/2)\Gamma(2-\mu)$; and $D_\mu$ is the intensity of the noise (see Eq.~(\ref{GCLT+A}) in the Appendix). The fact that we have preferred to work with the interval $1 < \mu < 2$ (i.e., $1/2 < s < 1$) is motivated by the understanding that the trajectory of an idealized test-particle performing a L\'evy flight must be a {\it continuous} curve (i.e., the Hausdorff dimension $d_f = \mu > 1$). The case $0 < \mu < 1$ is mathematically similar, however we do not consider it for classical particles. 

In order to understand the route to particle dynamics, let us reiterate on the above proposal that the noise term $F_\mu (t)$ is generated through nonlinear interactions between the coupled nonlinear oscillators in Eq.~(\ref{4s+}). That is, the nonlinear couplings in Eq.~(\ref{4s+}) (vibrations of the staircase jets) produce a noisy background in L-mode plasma, which is perceived by the plasma particles through the scattering on plasma waves. For $s < 1$ (nonlocal ordering present), these scattering processes could actually be very hard (the Langevin source is of the L\'evy type), leading to long-range jumps along the coordinate $n$ (see Appendix, Eq.~(\ref{Jump-l+A})), and to the L\'evy-fractional kinetic Eq.~(\ref{LFDiff}) for the probability density. Note that the quasilinear transport paradigm \cite{189,ZaslavskyUFN,Sagdeev} does not really apply here (if not for $s=1$ exactly). From this picture, the noise term in the Langevin Eq.~(\ref{Lvin}) is obtained directly from NLSE~(\ref{1}) by summing over $k$ the singular ``voices" from all oscillators in Eq.~(\ref{4s+}). This yields 
\begin{equation}
F_{\mu} (t) = \frac{\beta}{2} \Theta_{\rm wp}\Big[\sum_{k, m_1, m_2, m_3} V_{k, m_1, m_2, m_3} \sigma^{s}_{m_1} \sigma_{m_2}^{* s} \sigma_{m_3} +~c.\,c.\Big],
\label{LWn} % Eq.~(\ref{LWn})
\end{equation}
where $c.\,c.$ means complex conjugation (needed to ensure that we are dealing with real quantities), and $\Theta_{\rm wp}$ is a coefficient, which characterizes the efficiency of wave-particle interactions.
%This noise process will be a white L\'evy noise for $s < 1$, and a white Gaussian noise for $s=1$, provided just that the stochasticity condition in Eq.~(\ref{K-value}) holds (i.e., the wave packet is broad enough or the nonlinearity parameter is much greater than 1, i.e., $\beta \gg 1$). 
So the claim is that the sum in Eq.~(\ref{LWn}) produces a white L\'evy noise for $s < 1$, and a white Brownian noise for $s=1$, provided just that the stochasticity condition in Eq.~(\ref{K-value}) holds (i.e., the wave packet is broad enough) and that the system of coupled nonlinear oscillators~(\ref{4s+}) is found in vicinity of an attracting stable state (Sec.~III\,B).  % or the nonlinearity parameter is much greater than 1, i.e., $\beta \gg 1$; the index of the noise is $\mu = 2s$ and incorporates the Brownian case $\mu = 2$). 
%stochastic function, an uncorrelated white-noise process, if (i) $s \leq 1$, (ii) $K\gg 1$, and (iii) the wave packet in NLSE~(\ref{1}) is broad enough, so that the stochasticity condition in Eq.~(\ref{K-value}) holds. This stochastic function will be (i) a white L\'evy noise for $s < 1$, and (ii) a white Brownian noise for $s=1$, which is a special case of the backbone map in Eq.~(\ref{3ss}). 
In this way of thinking, we could trace the origin of L\'evy flights (and, more generally, the origin of nonlocal transport of particles in the radial direction) back to nonlinear interactions among the eigenstates in NLSE system~(\ref{1}). The subquadratic ($s < 1$) power nonlinearity is very important in this regard as it appears to be the wanted type of nonlinearity that leads directly to the white L\'evy noise (as an alternative to the Brownian noise), and to the Riesz derivative in the fractional diffusion Eq.~(\ref{LFDiff}). Note that the sum in Eq.~(\ref{LWn}) goes over {\it all} combinations of the wave numbers $k, m_1, m_2$, and $m_3$, and in this sense includes both the next-neighbor and far-distant couplings. Were the summation in Eq.~(\ref{LWn}) reduced to the next-neighbor couplings only, then a different description would result \cite{EPL} corresponding to a pseudochaotic dynamics \cite{Report,JMPB}.  

\subsubsection{Self-consistent model}

Turning, next, to a self-consistent setting, one must amend the diffusion equation~(\ref{LFDiff}) with a potential function due to the staircase transport barriers. This task has been already considered in the framework of the ``comb" model (see Ref. \cite{PRE18}) and leads to a self-consistent L\'evy-fractional Fokker-Planck equation (FFPE)  
\begin{equation}
\frac{\partial}{\partial t} f (n, t) =  \left[D_\mu \frac{\partial^\mu}{\partial |n|^\mu} + \frac{1}{\eta} \frac{\partial}{\partial n} \Phi^{\prime}(n)\right] f (n, t),
\label{FFPE} % Eq.~(\ref{FFPE})
\end{equation} 
where $\eta$ is the plasma viscosity; $\Phi^\prime (n) = d\Phi(n)/dn$; and $\Phi (n)$ is the comb potential mimicking the staircase structure (so $-\Phi^\prime (n)$ is the potential force along $n$ arising from the presence of the barriers; this force competes with the noiselike force $F_\mu (t)$ driving the L\'evy flights; see Fig.~3). 

\begin{figure}
\includegraphics[width=0.54\textwidth]{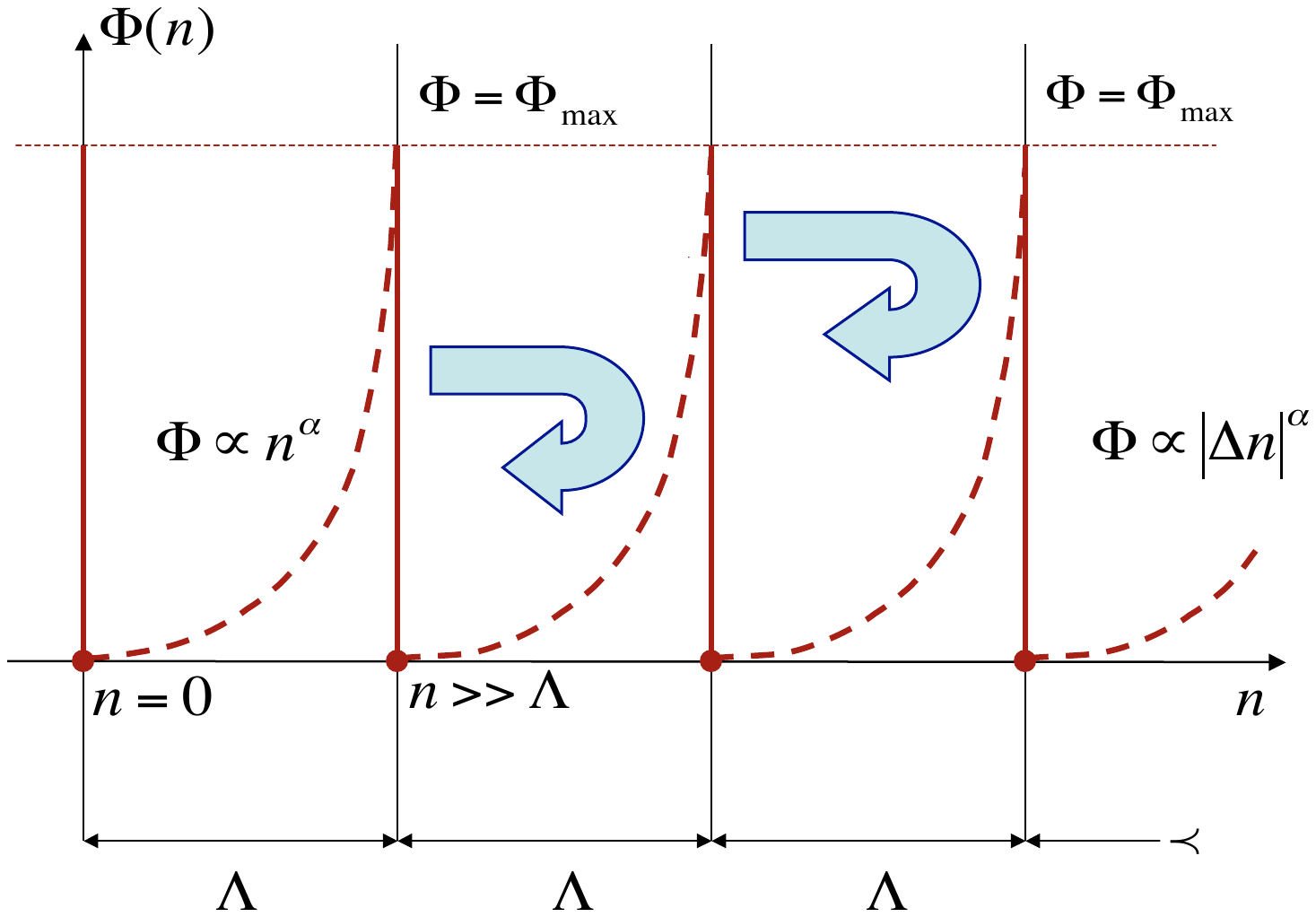}
\caption{\label{} The comb model of the plasma staircase. The potential function $\Phi(n)$ is modeled by a power-law dependence $\Phi(\Delta n) \propto |\Delta n|^\alpha$ at each step of the periodic staircase structure. In a simplified version of the model the distance between the steps, $\Lambda$, is assumed to be so long that one may extend the power-law dependence $\Phi(n) \propto n^\alpha$, good for $0 < n \lesssim \Lambda$, virtually to infinity. There is an upper threshold on $\Phi (n)$ above which the potential function does not affect the avalanches significantly, and this is set to be at $\Phi_{\max} = \Phi(\Lambda)$. We assume $\Phi_{\max}$ to be large compared with the characteristic kinetic energy of the avalanches. Similarly to Fig.~2, the U-turn arrows represent the avalanches confined between the staircase steps. Adapted from Ref. \cite{PRE18}.}
\end{figure}

The main idea behind the comb model is that the jet zonal flows and transport avalanches form a strongly coupled dynamical system and can exchange momentum, entropy and energy via the turbulent Reynolds stress. So the avalanches can naturally enhance or weaken the barriers, which would stabilize, then, at the edge of the localization-delocalization threshold. The edge property, in its turn, dictates a very specific shape to the $\Phi (n)$ dependence, which could be predicted at the delocalization threshold using general arguments \cite{PRE18}. According to the model, the avalanches are coherent structures, which grow to macroscopically noted sizes through the complex processes of mode coupling and the build-up of correlations (the reader might refer to e.g., the analysis of Refs. \cite{PLA2014,JPP}, where the occurrence of plasma avalanches was pursued starting from the paradigmatic Hasegawa-Wakatani model of electrostatic drift-wave turbulence \cite{Hasegawa,Horton}). While we do not discuss these processes here, we leave it to a remark that the origin of coherent structures has been (and still is) an important theme for tokamak plasma (e.g., Refs. \cite{JJR,Politzer,JJRPLA,Ippolito}; references therein). 

Somehow the avalanches can propagate radially at supersonic speeds (that is, much faster than the characteristic thermal velocities in the bulk plasma) \cite{Politzer,Ippolito}. As the avalanches can trap and convect particles, the resulting radial motion of these particles will appear as a sequence of almost instantaneous jumps in the radial direction \cite{PLA2014,JJR}. We associate these jumps with a L\'evy-flight motion along the coordinate $n$ and, mathematically, with the Riesz operator in the self-consistent FFPE~(\ref{FFPE}). 

\subsubsection{Taking a deep breath}

It is worth emphasizing that FFPE~(\ref{FFPE}) is formulated for the radial transport of particles, not waves. Indeed the transport of waves in a classical (not quantum \cite{PRE17,PRE19}) NLSE~(\ref{1}) is forbidden by the condition $s < 1$, and is only allowed for $s=1$ (above a nonzero threshold in $\beta$) \cite{PRE14,EPL}. The implication is that the plasma staircase is a structurally stable, robust dynamical system, in which the barriers (jet zonal flows) cannot diffuse away along the coordinate $n$ (unless quantum-tunneling effects are introduced \cite{PRE17,PRE19} or the driving noise process degenerates into a Brownian white noise in the limit $s\rightarrow 1$). It is in fact the nonlinear oscillations of the staircase jet zonal flows around their equilibrium positions that, according to our model, are sources of the white L\'evy noise in the medium. These self-consistently generated L\'evy noises act, in their turn, as a driving mechanism for the radial propagation of plasma particles on L\'evy flights, leading to the L\'evy-fractional diffusion Eq.~(\ref{LFDiff}). The fact that the staircase dynamical patterns generate just the L\'evy (not Brownian) noises is actually clear from the long-range correlated nature of the staircase self-organization (accounted for by the subquadratic power $s < 1$), and mathematically corresponds to the inclusion of the far-distant couplings in Eq.~(\ref{LWn}) (on an equal footing with the next-neighbor couplings).     

Regardless of the staircase implications, FFPE~(\ref{FFPE}) is a fundamental kinetic equation, which is obtained for many complex systems with long-range dependence under certain conditions (Refs. \cite{Klafter,Ch2007,Klafter2004,Report}; references therein). The competition between nonlocality, contained in the Riesz derivative~(\ref{Def+}), and nonlinearity, contained in the $\Phi (n)$ dependence, advances the important problem of confined L\'evy flight \cite{Klafter2004,Ch2004}, which had been a matter of some attraction in the literature \cite{PRE18,Ch2007}. In the Appendix to this paper, we present, for the reader's convenience, a derivation of FFPE~(\ref{FFPE}) for stochastic Markov chains in the long-wavelength limit (see Eq.~(\ref{FFPE+A}) and its hybrid generalization in Eq.~(\ref{FFPE+AA})). The derivation is aimed to demonstrate how fractional operators arise from the microscopic properties to dynamics, and uses the idea of complex transition probability in reciprocal space \cite{PLA2014} enabling a simple factorization procedure.    

\subsection{The issue of nonlocal transport}

Theoretically, the issue of nonlocal transport is a challenge because it violates the Fickian transport paradigm \cite{Fick,Sokolov} that fluxes at a point are decided by gradients at the same point. In magnetically confined fusion plasma, the interest in nonlocal transport was greatly fueled by the need to understand the behavior of cold pulses \cite{Pulse,Hariri} and the elusive ``uphill transport" \cite{Milligen04,Castillo06}, by which one means that, at some specific regions of the space, the flux may have the same direction as the gradient. Often the uphill transport is associated with the transport of voids propagating from the plasma edge inward, while the usual plasma avalanches propagate from the plasma core outward. Indeed the fractional flux is in general asymmetric and, for steady states, has a negative (toward the core) component that enhances confinement and a positive component that increases toward the edge and leads to poor confinement \cite{Castillo06}. In this work, we do not quite distinguish between the two components to the transport, and we tacitly assume that the results to be claimed for the ``avalanches" (e.g., event-size distribution, etc.) equally apply to the negative component as well. An evidence of nonlocal phenomena in tokamak plasma was provided by perturbative experiments \cite{Mantica,Mantica_etal} with plasma edge cooling and heating power modulation, indicating anomalously fast transport of edge cold pulses to plasma core, not compatible with major diffusive time scales. Recent progresses on experimental analysis and theoretical models of nonlocal transport applied to fusion plasma are reviewed in Ref. \cite{Ida}. 

\subsection{Adding white Brownian noise}
 
In writing FFPE~(\ref{FFPE}) we have assumed that the white L\'evy noise was the only stochastic driving process acting on plasma particles, and we associated the origin of this noise with nonlinear interactions between the staircase jet zonal flows$-$thought of as a system of coupled nonlinear oscillators in the backbone approximation. In a realistic jet plasma system, however, that won't be the only noise process to consider. For instance, at the small scales, one would naturally hear, in addition to the L\'evy noise in Eq.~(\ref{LWn}), also a white Brownian noise due to the Coulomb collisions and other internal frictional processes (e.g., quasilinear scatterings on the plasma waves, etc. \cite{Sagdeev,189}). If we were smart enough to include the Brownian noise already from the outset, we had to introduce, instead of FFPE~(\ref{FFPE}), a more general kinetic equation (see Appendix A for derivation)
\begin{equation}
\frac{\partial}{\partial t} f (n, t) =  \left[D_\mu \frac{\partial^\mu}{\partial |n|^\mu} + D \frac{\partial^2}{\partial n^2} + \frac{1}{\eta} \frac{\partial}{\partial n} \Phi^{\prime}(n)\right] f (n, t),
\label{FFPE+} % Eq.~(\ref{FFPE+})
\end{equation} 
where $D$ is the diffusion coefficient due to collisional processes and is defined as the intensity of the Brownian noise. Equation~(\ref{FFPE+}) is a hybrid kinetic equation, which contains the nonlocal (Riesz) and collisional (Brownian) diffusion on essentially an equal footing. We note in passing that the Brownian noise is the special case of the white L\'evy noise in the limit $\mu \rightarrow 2$, and it is the only L\'evy stable process to generate finite moments at all orders \cite{Report}. Setting $\mu\rightarrow 2$ in FFPE~(\ref{FFPE+}), one sees that the Riesz term vanishes in virtue of $\Gamma_\mu\rightarrow+\infty$, leaving solely the Brownian diffusion behind. 

In the applications of nonlocal transport (e.g., Refs. \cite{Pulse,Mantica}) it is convenient to think of the L\'evy noise as involving a nonlinear threshold condition in that the intensity $D_\mu$ is only nonzero above a certain critical value of the average gradient generating the instabilities and vanishes otherwise. Then FFPE~(\ref{FFPE+}) would readily switch between the local (e.g., collisional, as well as Gaussian quasilinear) transport in the parameter range of subcritical behavior, and nonlocal (L\'evy style) transport above the criticality. Note, in this regard, that FFPEs similar to Eq.~(\ref{FFPE+}) with a combination of L\'evy and ordinary diffusion have been also discussed in connection with the dynamics of protein fast-folding and the motion of excitations and proteins along polymer chains \cite{Ralf1,Ralf2}.

\section{Steady-state solution}

Next we look into a steady-state ($\partial / \partial t = 0$) solution of FFPE~(\ref{FFPE}). {\it A priori} we expect this solution to satisfy a boundary value problem that is (hopefully) consistent with the nonlocal nature of the Riesz operator in Eq.~(\ref{Def+}), mathematically a very nontrivial task \cite{Ch2007,Klafter2004}. The latter task, in its turn, is related with the exact analytic form for the potential function, $\Phi (n)$, which is staircase specific.  

\subsection{The model potential}

%It is understood that a particle (L\`evy walker) caught in a staircase will undergo a certain set of boundary conditions due to the periodic structure of the potential function, $\Phi(n)$. 

Aiming at a self-consistent description, we again refer ourselves to the comblike potential function $\Phi(n)$, which was introduced in Ref. \cite{PRE18} as a working model of the plasma staircase. The comb is built as a periodic sequence of spatially separated ``teeth," the spikes of $\Phi(n)$, which are reset after a spatial period, $\Lambda$, long compared with the collisional lengths (see Fig.~3). The teeth have tunable shape, and the shape effect of the staircase on the probability distribution of plasma avalanches has been addressed \cite{PRE18}. Here, we consider a simplified version of the comb model according to which the staircase steps are so long that one might formulate a transport problem between the neighboring teeth only (that is, one tooth at the origin and the next one virtually at infinity, where $\Phi (n)$ is reset). For this purpose, we assume that the particle is initiated at point $n=0$, where it finds a perfectly reflecting boundary corresponding to $\Phi(n) = +\infty$ for $n \leq 0$. Then for $n>0$ we postulate a power-law dependence $\Phi(n) = n^\alpha/\alpha$, where $\alpha$ is a power exponent. We consider $\alpha$ to be a free parameter, which characterizes the shape of the teeth within each staircase period. %The power-law dependence $\Phi(n) = n^\alpha/\alpha$ is truncated at the next step of the staircase (virtually, at infinity), where $\Phi(n)$ is reset to zero. 
A similar model but with a symmetric power-lawlike potential $\Phi(n) = |n|^c / c$ for $-\infty < n < +\infty$ was considered by Metzler {\it et al.} \cite{Ch2007}. In our case, because $\Phi^\prime (n) \equiv 0$ for all $n < 0$, there is a further simplification, making it possible to reduce the limits of improper integration in the Riesz fractional operator~(\ref{Def+}) to the positive semi-axis only. This yields a truncated operator in accordance with 
\begin{equation}
\frac{\partial^\mu}{\partial |n|^\mu} f (n) \Longrightarrow \frac{1}{\Gamma_\mu}\frac{\partial^2}{\partial n^2} \int_{0}^{+\infty}\frac{f (n^\prime)}{|n-n^\prime|^{\mu - 1}} dn^\prime,
\label{Def++} % Eq.~(\ref{Def++})
\end{equation}  
where we have also omitted the dependence over time in view of the steady-state condition $\partial f (n, t) / \partial t = 0$. We consider the truncated operator in Eq.~(\ref{Def++}) as representing the general form of the Riesz derivative in semi-infinite space $n \geq 0$, with a perfectly reflecting left boundary satisfying $\Phi(n) = +\infty$ for $n \leq 0$ and $\Phi^\prime (n) \equiv 0$ for $n < 0$.

\subsection{Steady-state solution for large $n$}

For large $n$, we can neglect the Brownian diffusion term compared to the L\'evy-flight term in FFPE~(\ref{FFPE+}), hence base our analysis on the fractional Fokker-Planck equation~(\ref{FFPE}), with $\partial f (n, t) / \partial t = 0$. Considering the boundary condition $\Phi^\prime (n) \equiv 0$ for $n < 0$, we have, with the aid of Eq.~(\ref{Def++}), 
\begin{equation}
- \frac{1}{\eta} \frac{\partial}{\partial n} (\Phi^{\prime}(n) f (n)) = \frac{D_\mu}{\Gamma_\mu} \frac{\partial^2}{\partial n^2} \int_{0}^{+\infty}\frac{f (n^\prime)}{|n-n^\prime|^{\mu - 1}} dn^\prime,
\label{VNL} % Eq.~(\ref{VNL})
\end{equation}  
where $\Phi^\prime (n) = n^{\alpha - 1}$ for $n > 0$. To enjoy smooth analytic dependence for $n\rightarrow +0$, it is required that the power exponent $\alpha > 2$. It is therefore ensured that the vanishing first derivative $\Phi^\prime (n) \rightarrow 0$ at the origin $n\rightarrow +0$ matches the condition $\Phi^\prime (n) \equiv 0$ on the negative semi-axis. 

To assess the solution for $n\rightarrow+\infty$, one might proceed as follows. If the L\'evy index is in the range $1 < \mu < 2$, then for large space lags $|n - n^\prime|\rightarrow+\infty$ one dares say that the inverse power law $1/|n - n^\prime|^{\mu - 1}$ is a slowly decaying function of the wave number, so one ventures to take it out of the integral sign, and to replace by a scaling factor $1/n^{\mu - 1}$ instead, after which the remaining improper integral over $n^\prime$ of the probability density is claimed to be equal to 1 by the conservation of the probability, i.e., $\int_0^{+\infty} f (n^\prime) dn^\prime = 1$. The result is that the right-hand side of Eq.~(\ref{VNL}) scales with $n$ as $1/n^{\mu + 1}$ (for asymptotically large $n$), typical for L\`evy distributions. Utilizing $\Phi^\prime (n) = n^{\alpha - 1}$ on the left-hand side, from Eq.~(\ref{VNL}) one arrives at the asymptotic power-law behavior
\begin{equation}
f(n) \simeq (\eta D_\mu / \Gamma_\mu)\, n^{-(\alpha+\mu - 1)},
\label{Tail} % Eq.~(\ref{Tail})
\end{equation}    
where $n\rightarrow+\infty$. It is clear from Eq.~(\ref{VNL}) that the steady-state $f(n)$ for large $n$ is defined by a competition between the nonlocality, contained in the Riesz fractional derivative~(\ref{Def++}), and the nonlinearity, contained in the $\Phi(n)$ dependence. If the L\'evy index approaches its Gaussian value, i.e., $\mu\rightarrow 2$, then the normalization parameter $\Gamma_\mu = - 2\cos(\pi\mu/2)\Gamma(2-\mu)$ goes to infinity (because the gamma function $\Gamma (2-\mu)$ diverges in this limit). The implication is that the power-law dependence in Eq.~(\ref{Tail}) would only occur for the fractional $\mu$ values for which the dynamics are nonlocal, otherwise the Brownian term in FFPE~(\ref{FFPE}) must be reinstalled. % For $\mu\rightarrow 2$, the local properties are reinstalled leading to a Gaussian random-walk operator in FFPE~(\ref{FFPE}).  

\subsection{Steady-state solution for small $n$}

Next we turn to the opposite limiting case of small $n$. In this regime, the Riesz term in FFPE~(\ref{FFPE+}) is much weaker than the Brownian term, suggesting that the steady-state solution near the origin ($n\rightarrow +0$) could be obtained by balancing the Brownian term to the nonlinear term. That would yield, similarly to Eq.~(\ref{VNL}),  
\begin{equation}
- \frac{1}{\eta} \frac{\partial}{\partial n} (\Phi^{\prime}(n) f (n)) = D \frac{\partial^2}{\partial n^2} f(n),
\label{VNL+} % Eq.~(\ref{VNL+})
\end{equation}  
from which a compressed ($\alpha > 2$) Gaussian %(in view of $\alpha \geq 2$) 
bell function 
\begin{equation}
f(n) \simeq f_0 \exp (-n^\alpha / \alpha\eta D)
\label{Comp} % Eq.~(\ref{Comp})
\end{equation} 
can be inferred. This compressed Gaussian behavior being analytically very appealing might yet be unsatisfactory for practical applications in that it does not take into account the possible sources and sinks of particles in the bulk of the staircase (not included in FFPE~(\ref{FFPE+})). 

To remedy, let us consider a more general situation, according to which the sources are concentrated at the origin $n=0$ and have the form of the Dirac's delta-pulse, i.e., $\hat S_+ [f (n)] = \delta (n)$, while the sinks are proportional to the probability density and are given by the dependence $\hat S_- [f (n)]=-\varepsilon f(n)$, where $\varepsilon$ is a coefficient that characterizes the sinks. We associate this dependence with the stabilizing effect of the shear flows on radial transport \cite{Itoh,Itoh2}. Then for $n > 0$, $n\rightarrow +0$, we may argue that the shape of the $f(n)$ distribution is defined by a competition between the sink term $\hat S_- [f (n)]=-\varepsilon f(n)$ and the Gaussian diffusion term, leading to  
\begin{equation}
- D \frac{\partial^2}{\partial n^2} f (n) = \hat{S}_{-}[f (n)] = -\varepsilon f(n), 
\label{StSt} % Eq.~(\ref{StSt})
\end{equation} 
from which a simple exponential decay of $f(n)$ would result, i.e., 
\begin{equation}
f(n) \simeq \exp (-\sqrt{\varepsilon/D}\,n). 
\label{Core} % Eq.~(\ref{Core})
\end{equation} 
The exponential function in Eq.~(\ref{Core}) introduces a characteristic scale into the transport model, i.e., $\lambda \simeq \sqrt{D/\varepsilon}$. Note that the probability density in Eq.~(\ref{Core}) decays much slower than the compressed Gaussian distribution in Eq.~(\ref{Comp}). The implication is that the sink term disperses the particles over much broader a region than the balance between the diffusion and the potential-function term in FFPE~(\ref{FFPE+}) would predict. In this regard, we consider the length scale $\lambda$ as a crossover scale from local (Gaussian) diffusion to nonlocal transport. 

\subsection{Orderings and lengths} 

Once the crossover scale is introduced, one might naturally argue that the scale-free distribution in Eq.~(\ref{Tail}) applies, if $n \gg \lambda$. The latter condition offers a quantitative measure of how ``large" are the large $n$'s, considered in Sec.~IV\,B. For the reasons of formal ordering, we must also require that the parameter $\lambda$ be much smaller than the spacing $\Lambda$ between the consecutive teeth of the staircase (see Figs.~2c and~3), that is, the crossover to nonlocal transport must occur within one staircase period. By adopting a hypothesis that the $\bf E \times \bf B$ staircase develops though the self-organization of vortical flows in magnetized plasma, we evaluate $\Lambda$ as the Rhines length for electrostatic drift-wave turbulence, i.e., $\Lambda \simeq \Lambda_{\rm Rh}$, where $\Lambda_{\rm Rh}$ designates the spatial scale separating vortex motion from drift-wavelike motion \cite{Naulin}, similarly to its fluid analog \cite{McIntyre}. Note that $\Lambda_{\rm Rh}$ scales as a square root of the $\bf E \times \bf B$ drift, leading to $\Lambda_{\rm Rh} \simeq \sqrt{u_E} \simeq \sqrt{E/B}$, where $u_{E} = |{\bf E} \times {\bf B}| / B^2$ is the drift velocity, $E$ is the radial electric field, and $B$ is the toroidal magnetic field. Using $\lambda \simeq \sqrt{D/\varepsilon}$, from the condition $\lambda \ll \Lambda \simeq \Lambda_{\rm Rh}$ we have $E \gg DB/\varepsilon$, that is, the staircase is a thresholded phenomenon, which may only occur if the radial electric field due to the drift waves exceeds a certain critical value (of the order of $DB/\varepsilon$). Note that the threshold appears to be higher in tokamaks with greater toroidal field $B$ and for plasmas with a lower $\varepsilon$ value.

\subsection{Weak localization of avalanches} 

The asymptotic probability density in Eq.~(\ref{Tail}) is a starting point to analyze the localization properties of plasma avalanches. First of all, it is clear that we are dealing with a localization phenomenon that is different from the familiar, strong localization on an exponentially fast drop-off \cite{And,Chacra} (because the decay of the $f(n)$ function is power-lawlike for $n\rightarrow+\infty$, implying the significant likelihood of under-barrier crossing). Following Ref. \cite{PRE18}, we refer to this type of localization phenomenon as ``weak" localization. Second, we distinguish between the behaviors with respectively finite and infinite second moments, and we associate the weak localization with a type of behavior when the second moments are {\it finite} \cite{Ch2007,Klafter}. Mathematically, that means that the integral $\int^n {n^\prime}^2 f (n^\prime) dn^\prime < +\infty$ must converge at infinity. This is possible, if and only if the steady-state $f(n)$ decays faster than an inverse-cubic drop-off, that is, faster than $\propto 1/n^{3}$ for $n\rightarrow+\infty$. A condition for that is that the power exponent in Eq.~(\ref{Tail}) is greater than 3, i.e.,   
\begin{equation}
\alpha+\mu - 1 > 3,
\label{Cond} % Eq.~(\ref{Cond})
\end{equation}    
leading to $\alpha > 4-\mu$. One sees that a L\'evy flyer in a free space cannot be weakly localized (because in the absence of potential fields the $\alpha$ value is zero; then the inequality in Eq.~(\ref{Cond}) says that the L\'evy index $\mu$ must be greater than 4 at odds with the L\'evy-Gnedenko generalized central limit theorem \cite{Gnedenko}). In the same spirit, a potential field $\Phi(n) \propto n^\alpha$ can weakly localize the avalanches, if (and only if) it grows sharply enough with $n$ for $n\rightarrow+\infty$, i.e., if the $\alpha$ value is greater than $\alpha_{\min} = 4-\mu$. In particular, the so-called Cauchy flights \cite{Mainardi}$-$characterized by $\mu \rightarrow 1$ and representing ballistic transport in radial direction$-$could be weakly localized, if $\alpha > \alpha_{\min} = 3$. If the $\alpha$ value is fixed, then the weak localization occurs for a family of L\'evy-flight processes obeying $\mu > 4-\alpha$ and $1 < \mu < 2$, which might or might not be satisfied.

\section{Event-size distribution of avalanches}

We have seen in the above that the analytical structure of $f(n)$ is divided between the core (Eq.~(\ref{Core})) and tail (Eq.~(\ref{Tail})) regions. The crossover between the two regions is at $\lambda \simeq \sqrt{D/\varepsilon}$ and is assumed to be small compared to the distance between the jets, i.e., $\lambda\ll \Lambda_{\rm Rh}$. With a permission that $\Lambda_{\rm Rh}$ is virtually infinite, i.e., the drift-wave turbulence is actually very strong, we might rely on the inverse power-law behavior in Eq.~(\ref{Tail}) for all $\lambda \ll n \lesssim \Lambda_{\rm Rh} \simeq +\infty$. The size distribution of avalanches is obtained as the probability for the random walker to {\it not} be dispersed by the Fokker-Planck dynamics after $\Delta n$ space steps in radial direction, motivating
\begin{equation}
w_\tau (\Delta n) = \left[\int_{0}^{+\infty} - \int_{0}^{\Delta n}\right] f (n^\prime)dn^\prime = \int_{\Delta n}^{+\infty} f (n^\prime)dn^\prime.
\label{Size} % Eq.~(\ref{Size})
\end{equation} 
It is understood that the avalanches are coherent structures, which means that the integration in Eq.~(\ref{Size}) produces the number density of such avalanches with sizes between $\Delta n$ and $\Delta n + d\Delta n$ (and not yet the number of the avalanches with sizes from $\Delta n$ to $+\infty$, which would be the integral $\int_{\Delta n}^{+\infty}w_\tau (\Delta n^\prime)d\Delta n^\prime$). Setting $\Delta n$ to be in the tail region, i.e., $\Delta n \gg \lambda$, with the aid of Eq.~(\ref{Tail}) one obtains
\begin{equation}
w_\tau (\Delta n) \propto (\eta D_\mu / \Gamma_\mu)\, \Delta n^{-\tau},
\label{Tail+} % Eq.~(\ref{Tail+})
\end{equation}    
where
\begin{equation}
\tau = \alpha + \mu - 2
\label{Tail++} % Eq.~(\ref{Tail++})
\end{equation}   
is the exponent of the power law. In the above we have promoted the coefficient $1 / \Gamma_\mu$ to emphasize that there is no asymptotic power-law behavior in the Gaussian limit, $\mu\rightarrow 2$. By examining Eq.~(\ref{Tail++}) one sees that the exponent $\tau$ increases with the increasing both $\alpha$ (the shape of the potential function) and $\mu - 2 < 0$ (distance to the Gaussian limit). That is, a stronger $\Phi(n)$ dependence, weaker nonlocal features would result, as a rule, in a steeper line of decay of the ensuing $w_\tau (\Delta n)$ distribution, with a narrower room for large-amplitude and extreme events. The latter conclusion, though, should be taken with a grain of salt in that a sharper $\Phi(n)$ can by itself act as a source of the free energy driving the avalanches (Sec.~V\,A.) Note that the exponent $\tau$ is always positive in view of $\alpha > 2$.

\subsection{The $\tau$ exponent and the state of marginal stability}

Let us now obtain the exponent $\tau$ self-consistently. For this, we shall arguably assume that the steady state of the staircase is stable and self-organized. The staircase being stable means it had occurred at a local free-energy minimum. In plasmas, the free energy is contained in temperature and pressure gradients that are sources of various instabilities \cite{189,Horton}. On the other hand, in driven systems, a class to which (as we argue) belongs the plasma staircase, the instabilities will be excited naturally thanks to the free-energy input, then the stability condition says that the local free-energy minimum is found at the edge of the instability drive, provided that the rate of the driving is (i) so slow that it allows the staircase system to self-organize, yet is (ii) strong enough to feed the staircase dynamical patterning against dissipation. This observation suggests that the shape of the potential function at the steady state (to be associated with the state of marginal stability of the staircase) is such that the $\Phi (n)$ dependence (i) localizes (weakly) any type of avalanche, with the L\'evy index $\mu$ in the interval $1 < \mu < 2$, and (ii) the localization of the Cauchy flights, with $\mu \rightarrow 1$, is at the edge of delocalization. With the aid of $\alpha_{\min} = 4-\mu$ this yields $\alpha = 3$ exactly at the marginality. Using Eq.~(\ref{Tail++}), we have $\tau = 2s +1$, where $\mu = 2s$ has been considered. 

The fact that there exists a marginally stable state that attracts the nonlinear staircase dynamics finds further support in the analysis of the avalanche-zonal flow coupling. In fact, assume that the function $\Phi(n)$ is perturbed such that it grows with coordinate $n$ faster than the least necessary to confine the avalanches (in our case faster than the cubic power of the coordinate). That means that there is excess free energy in the jet flows that is contained in the increased spatial gradient, $\Phi^\prime (n)$. Then the system seeking a new stable state will release the excess energy by exciting secondary instabilities in the plasma. Part of these instabilities will be absorbed by the avalanches through the processes of mode coupling and the build-up of correlation \cite{JJR,JJRPLA,JPP}, the result being that the avalanches grow, and their magnitudes increase. Simultaneously, the height of the barriers containing the avalanches will be lowered due to energy conservation$-$a process that would naturally flatten the $\Phi(n)$ dependence. If the lowering is significant enough, then the avalanches may escape the barrier, and this will take away some energy from the staircase in the form of radial kinetic energy. To this end, the $\Phi(n)$ function becomes too flat to confine the avalanches effectively, while the radial transport has intensified. Nonlinearly, the intensification of radial transport will act as to reconstruction the barriers (because the newly formed avalanches will transport azimuthal momentum up the gradient of the azimuthal flow, hence will drive the zonal-flow shear while moving outwards). The phenomenon had been seen directly in the JET experiments \cite{Xu}. During this process, energy is transferred from the avalanches to the zonal flows via the turbulent Reynolds stress, resulting in suppression of fluctuations in-between the staircase steps and simultaneous ensuing intensification of the jet poloidal flows. 

The asymptotic dynamical state of the system crucially depends on the rate of external forcing \cite{PLA2014,JPP}: If the free-energy input rate is slow enough, then the system eventually returns in the vicinity of the marginally stable state from which it had departed. When this occurs, we are said that the initial $\Phi(n)$ has been successfully reinstalled. On the contrary, if the driving is strong in a sense, i.e., the free-energy input rate goes above a certain nonlinear threshold, then the driven system would end up in ever-continued predator-preylike oscillations around the marginally stable state, a type of dynamics characterized by periodic weakening of the barriers due to their cross-talk with the radial transport. This type of behavior has been reported for magnetically confined fusion plasma by Schmitz {\it et al.} \cite{Schmitz}. 

In what follows, we assume that the rate of the driving is so slow that any eventual deviations from the marginally stable state dissipate before they could trigger the predator-preylike oscillations. This said, we refer to the value $\alpha = 3$ as the natural stability limit for the plasma staircase that attracts the nonlinear avalanche-jet zonal flow dynamics. Note that the assumption of weak driving is crucial in this regard: It is due to this assumption that one may rely on the power-law reduced event-size distribution in Eq.~(\ref{Tail+}) and the associated multi-scale dynamical properties. Should the assumption of slowness of the driving be invalidated, a characteristic scale will be introduced through the excitation of the predator-preylike oscillations, while the multi-scale features to the dynamics would generally be suppressed. 

One sees that by accepting the power-law in Eq.~(\ref{Tail+}) one must also require that the system is driven ``slowly" enough in that it is allowed to produce complex and multi-scale features {\it naturally} through the self-organization into a state of marginal stability, otherwise it undergoes forced nonlinear oscillations in which the turbulence preys on self-organized transport barriers \cite{Itoh,Itoh2,Schmitz}. Theoretically, a crossover from multi-scale to auto-oscillatory dynamics has been discussed for driven systems in Refs. \cite{NJP,Chapter}, where one also finds a general condition on the limit driving rate.

\subsection{Finding the $s$ value}

Our next (and final) task is to assess the $s$ value at the marginality. For this, let us observe, following the analysis of Refs. \cite{PRE17,EPL}, that the nonlinear frequency shift $\Delta\omega_{\rm NL} = \beta |\psi_n|^{2s} \simeq \beta / (\Delta n)^s$ in NLSE~(\ref{1}) has the sense of the effective ``temperature" of nonlinear interaction. In a thermodynamically stable state, this temperature controls the reservoir of the free energy driving the instabilities. Denoting the free energy as $\mathcal{E}$ and the temperature as $T$, we have the ordering $\mathcal{E} \simeq T$ near the stable state. On the other hand, the total free energy, $\mathcal{E}$, may be represented as the energy density, $\varrho$, times the system's volume, $\Omega$, i.e., $\mathcal{E} \simeq \varrho \Omega$. The energy density, $\varrho$, is none other that the thermodynamic pressure, $P$, whereas the volume $\Omega$ is expressed, for the NLSE system, as the number of states, i.e., $\Omega \simeq \Delta n$. Putting all the various pieces together, we have $T \simeq \Delta\omega_{\rm NL} = \beta / (\Delta n)^s \simeq \varrho \Omega \simeq P\Delta n$, from which the familiar polytropic equation of state \cite{Levich} 
\begin{equation}
P\Omega^{\gamma} \simeq \beta = {\rm const}
\label{Adi} % Eq.~(\ref{Adi})
\end{equation}   
can be deduced, with $\gamma = s+1$. In a basic theory of ideal gases, the polytropic Eq.~(\ref{Adi}) describes an adiabatic (no entropy generation) process. The adiabatic character means that there is no heat and energy exchange with the exterior. In this regard, the interpretation of the parameter $\gamma$ as ``adiabatic" index finds its justification in the two conserved quantities of the NLSE model the Hamiltonian (Eqs.~(\ref{6}) and~(\ref{6s+})) and the total probability (Eq.~(\ref{TP})). When applied to the plasma staircase as driven system, the adiabatic character would imply that the driving rate is infinitesimally slow, such that the system is allowed to accommodate any free-energy input before a new portion of the free energy is again introduced. One sees that adiabaticity is a very good property in that it guarantees both the polytropic form of Eq.~(\ref{Adi}) and the power-law event-size distribution in Eq.~(\ref{Tail+}). 

As is well known from classic thermodynamics of ideal gases, the adiabatic index has the following general representation \cite{Levich}: $\gamma = (2+\zeta)/\zeta$, where $\zeta$ is the number of degrees of freedom of a molecule, so that for, for instance, a monoatomic gas, with three degrees of freedom, $\zeta = 3$ and $\gamma= 5/3$. Combining with $\gamma = s+1$, one sees that $s = 2/\zeta$. If $\zeta\rightarrow\infty$, then $s\rightarrow +0$, hence NLSE~(\ref{1}) reduces to a linear Schr\"odinger equation in which the energy levels are all shifted by the same value of $\beta$, i.e., $E_n \longrightarrow E_n + \beta$. For the nonlinear model with self-interference among the waves, the number of degrees of freedom is obtained self-consistently from the Langevin Eq.~(\ref{4s+}) as the sum of the power exponents standing for the stochastic term, i.e., $\zeta = 2s + 1$. Equating this to $\zeta = 2/s$, one arrives at a simple quadratic equation for $s$, i.e., 
\begin{equation}
2s^2 + s - 2 = 0,
\label{QEq} % Eq.~(\ref{Qeq})
\end{equation}   
yielding $s = ({\sqrt{17}} - 1)/{4} \simeq 0.78$. Once the value of $s$ is known, one takes stock of the entire pool of the parameters introduced above, i.e., $\mu = 2s$; $\zeta = 2/s$; $\gamma = s+1$; and $\tau = 2s+1$. Note that the $\tau$ value is obtained using Eq.~(\ref{Tail++}) and the result that $\alpha = 3$ at the free energy minimum. We have collected our findings in Tables~I and II.  

The fact that there exists a nontrivial $s$ value satisfying the complex bargain between the various parameters and processes involved, i.e., $s = ({\sqrt{17}} - 1)/{4} \simeq 0.78$, shows that the coupled avalanche-jet zonal flow system can, in fact, stabilize itself in the vicinity of a dynamical steady state, in which it would release excess free energy in the form of avalanches with a broad event-size distribution, i.e., $w_\tau (\Delta n) \propto \Delta n^{-\tau}$ for $\Delta n \gg 1$. The $\tau$ exponent is found to be $\tau = ({\sqrt{17}} + 1)/{2} \simeq 2.56$ and is sensitive to both the $s$ value and the condition that the system as a whole is near its free energy minimum, i.e., $\alpha = 3$. We consider this nonequilibrium steady state as attracting the nonlinear staircase dynamics. Note that the avalanches prove to be {\it nonlocal} transport events, which is immediately seen from the result that the L\'evy index $\mu$ is smaller than 2, i.e., $\mu = ({\sqrt{17}} - 1)/{2} \simeq 1.56$. Note, also, that the polytropic exponent $\gamma = ({\sqrt{17}} + 3)/{4} \simeq 1.78$ appears to be remarkably close to (although slightly larger than) the paradigmatic value $\gamma = 5/3 \simeq 1.67$ for monoatomic ideal gas \cite{Levich}. The hint is that the system of coupled nonlinear oscillators in Eq.~(\ref{4s+}) behaves as well as it was an ideal ``monoatomic" gas embedded in a hypothetical fractal embedding space, with a fractional number of the embedding dimensions being equal to $\zeta = ({\sqrt{17}} + 1)/{2} \simeq 2.56 < 3$. A mathematical case of such hypothetical spaces has been discussed in Ref. \cite{PRE97} based on the notion of fractal manifold. 

In the applications of statistical mechanics of complex systems one often writes the $\tau$ exponent in Eq.~(\ref{Tail+}) as $\tau = (1+\kappa)/\kappa$, where $\kappa$ is the so-called ``kappa" parameter and is introduced to interpolate between the statistical distributions in the core and tail regions (e.g., Ref. \cite{NG}; references therein). With the aid of $\tau = 2s + 1$ one gets $\kappa = 1/2s = 1/\mu$. One sees that the kappa value is none other than the inverse L\'evy index $\mu$ and is the relevant parameter to characterize the nonlocal transport by plasma avalanches (as much as the $\mu$ value). Using for the exponent $s$ its self-consistent value $s = ({\sqrt{17}} - 1)/{4} \simeq 0.78$ at the steady state, one also obtains $\kappa = ({\sqrt{17}} + 1)/8 \simeq 0.64$. %We shall be back to this finding in Sec. IV, where results from numerical simulations of the dynamic staircase system are presented.  

\begin{table}[b]
\caption{A summary of the exponents that have been used to characterize the coupled avalanche-jet zonal flow system and the ensuing statistical distributions of plasma avalanches.}

\label{tab:1}       % Give a unique label
%
% Follow this input for your own table layout
%
\begin{tabular}{p{1.1cm}p{2.9cm}p{4.2cm}}
\hline
\hline
Value & Defined through & Description \\
\hline
$s$& $|\psi_n|^{2s} \equiv (|\psi_n|^2)^{s}$ & subquadratic power exponent\\
$\mu$& $F_\mu (t)$, Eq.~(\ref{GCLT+A}) & the index of L\'evy stable noise\\
$\alpha$& $\Phi(\Delta n) \propto |\Delta n|^\alpha$ & potential-function exponent\\
$\tau$& $w_\tau (\Delta n) \propto \Delta n^{-\tau}$ & event-size distribution\\
$\gamma$& $P\Omega^{\gamma} \simeq \beta = {\rm const}$ & polytropic exponent\\
$\zeta$& $\gamma = (2+\zeta)/\zeta$ & number of degrees of freedom\\
$\kappa$& $\tau = (1+\kappa)/\kappa$ & kappa exponent\\
\hline
\hline
\end{tabular}
%$^a$ Table foot note (with superscript)
\end{table}
% 
%

\begin{table}[b]
\caption{Self-consistent values of the exponents $s$, $\mu$, $\alpha$, $\tau$, $\gamma$, $\zeta$, and $\kappa$ in the vicinity of the steady state. Second column: analytic representations in terms of the subquadratic power exponent $s$. Third column: exact values using transcendental numbers. Fourth column: approximate numerical estimates rounded to two decimal places (marked by the $^\pm$ sign).}

\label{tab:1}       % Give a unique label
%
% Follow this input for your own table layout
%
\begin{tabular}{p{1.1cm}p{3.1cm}p{2.4cm}p{1.5cm}}
\hline
\hline
Value & Math expression & Exact result & Approx\\
\hline
$s$& $|\psi_n|^{2s} \equiv (|\psi_n|^2)^{s}$ &$({\sqrt{17}} - 1)/{4}$& $0.78^\pm$\\
$\mu$& $\mu = 2s$ &$({\sqrt{17}} - 1)/{2}$& $1.56^\pm$\\
$\alpha$& $\alpha > 4-\mu$ &$3$& $3$\\
$\tau$& $\tau = 2s + 1$ & $({\sqrt{17}} + 1)/{2}$ & $2.56^\pm$\\
$\gamma$& $\gamma = s+1$ & $({\sqrt{17}} + 3)/{4}$ & $1.78^\pm$\\
$\zeta$& $\zeta = 2s + 1 = 2/s$ & $({\sqrt{17}} + 1)/{2}$ & $2.56^\pm$\\
$\kappa$& $\kappa = 1/2s$ & ${\sqrt{17}} + 1)/8$ & $0.64^\pm$\\
\hline
\hline
\end{tabular}
%$^a$ Table foot note (with superscript)
\end{table}

\subsection{Fitting to a Fr\'echet distribution}

In order to validate the theory prediction that the event-size distribution of plasma avalanches has a fat power-law tail, we have compared the $w_\tau (\Delta n)$ dependence in Eq.~(\ref{Tail+}) with the corresponding numerical dependencies obtained from gyrokinetic simulations \cite{DF2017,Horn2017} of the plasma staircase using flux-driven gyrokinetic code \textsc{Gysela} \cite{Sarazin}. The parameters of the simulation were those mimicking the Tore Supra shot $\#$ 45 511 \cite{Shot}: same parameters were used in Ref. \cite{DF2015} to generate the staircase pattern reproduced in Fig.~1; as well as those mimicking the consequent shots $\#$ 47 670 and $\#$ 47 923, performed under similar plasma conditions. A close fit to the probability density function was obtained using a Fr\'echet distribution (e.g., Ref. \cite{Fresh}) 
\begin{equation}
{\mathcal{F}}_\kappa (\Delta n) = \frac{(G (\Delta n))^{1+\kappa}}{a} \exp [-G (\Delta n)]
\label{Fresh-G} % Eq.~(\ref{Fresh-G})
\end{equation}   
with
\begin{equation}
G (\Delta n) = \left(1 + \kappa \frac{\Delta n - b}{a} \right)^{-1/\kappa}.
\label{Fresh-G+} % Eq.~(\ref{Fresh-G+})
\end{equation} 
The Fr\'echet distribution belongs to a family of the generalized extreme value distributions \cite{Haan} and is a relevant tool to model the maxima of finite sequences of random variables. In the above $a$ and $b$ are, respectively, the scale and the location parameters and were fitted in the simulation to the actual positions and strengths of the avalanches, so this gave $a \simeq 10$ and $b \simeq 44$; and $\kappa$ is the ``kappa" parameter, which fitted the shape of the distribution in the entire range of the $\Delta n$ variation. For $\Delta n \gtrsim 40$, a behavior compatible with the power law 
\begin{equation}
{\mathcal{F}}_\kappa (\Delta n) \sim \Delta n^{-(1+\kappa)/\kappa}
\label{Fresh_P} % Eq.~(\ref{Fresh_P})
\end{equation}  
was recognized, with the optimum fit at $\kappa \simeq 0.67$ in good agreement with the NLSE result $\kappa = ({\sqrt{17}} + 1)/8 \simeq 0.64$ (see Table~2). %This suggests that in terms of the plasma staircase we are really dealing with a system that generates extreme avalanches and power laws. 

To illustrate these findings, we have plotted in Fig.~4 the \textsc{Gysela} computed probability density in comparison with the Fr\'echet distributions for the different $\kappa$ values (min $\kappa = 0.501$; max $\kappa = 0.8$). Also in Fig.~4 we have plotted the cumulative root-mean-square error between the numerically computed (\textsc{Gysela}) and template (Fr\'echet) distributions, showing that the cumulative error tends to saturate above $\Delta n \sim 1\cdot 10^2$. These numerical plots are further reprocessed in Fig.~5, where we have singled out the Fr\'echet distribution for the very specific value $\kappa \simeq 0.67$ for which the optimum fit was obtained. On the right panel of Fig.~5 we displayed the normalized root-mean-square error in a percentile to the maximum error at $\kappa \simeq 0.85$ (actually the max $\kappa$ value analyzed in the simulations), from which it is clear that the fitting quality is maximized for $\kappa \simeq 0.67$. Complementing the numerical fits is a coarse experimental distribution of the probability density (the steplike function (gray color); shown in a discrete form in Fig.~4, and in an interpolated form in Fig.~5) obtained by concatenating 15 different occurrences of staircase observations in well-diagnosed Tore Supra shots $\#$ 47 670 and $\#$ 47 923. One sees, from this distribution, that the experimental diagnostics alone is actually too rough to conclude the $\kappa$ value with wanted precision (and it was not, in fact, our goal to perform this task), yet we were able to reconstruct, using the \textsc{Gysela} code, a virtual reality incorporating the experimental distribution to a very good extent, with a synthetic statistics actually covering the entire range of the $\Delta n$ variation.

Finally, we acknowledge the limitations due to the finite system-size, yet we confirm that the data had spanned a reasonably wide spatial segment (in practical terms, up to 15 single avalanche sizes), enabling a statistical fit to the power-law in Eq.~(\ref{Fresh_P}). Numerically, the occurrence of the power-law subrange in Eq.~(\ref{Fresh_P}) was confirmed for $\Delta n$ being approximately four times larger than the turbulence auto-correlation length, which corresponds with the typical influence range of the meso-scale self-organization \cite{DF2017,Horn2017}.

\begin{figure}[t]
\includegraphics[width=0.49\textwidth]{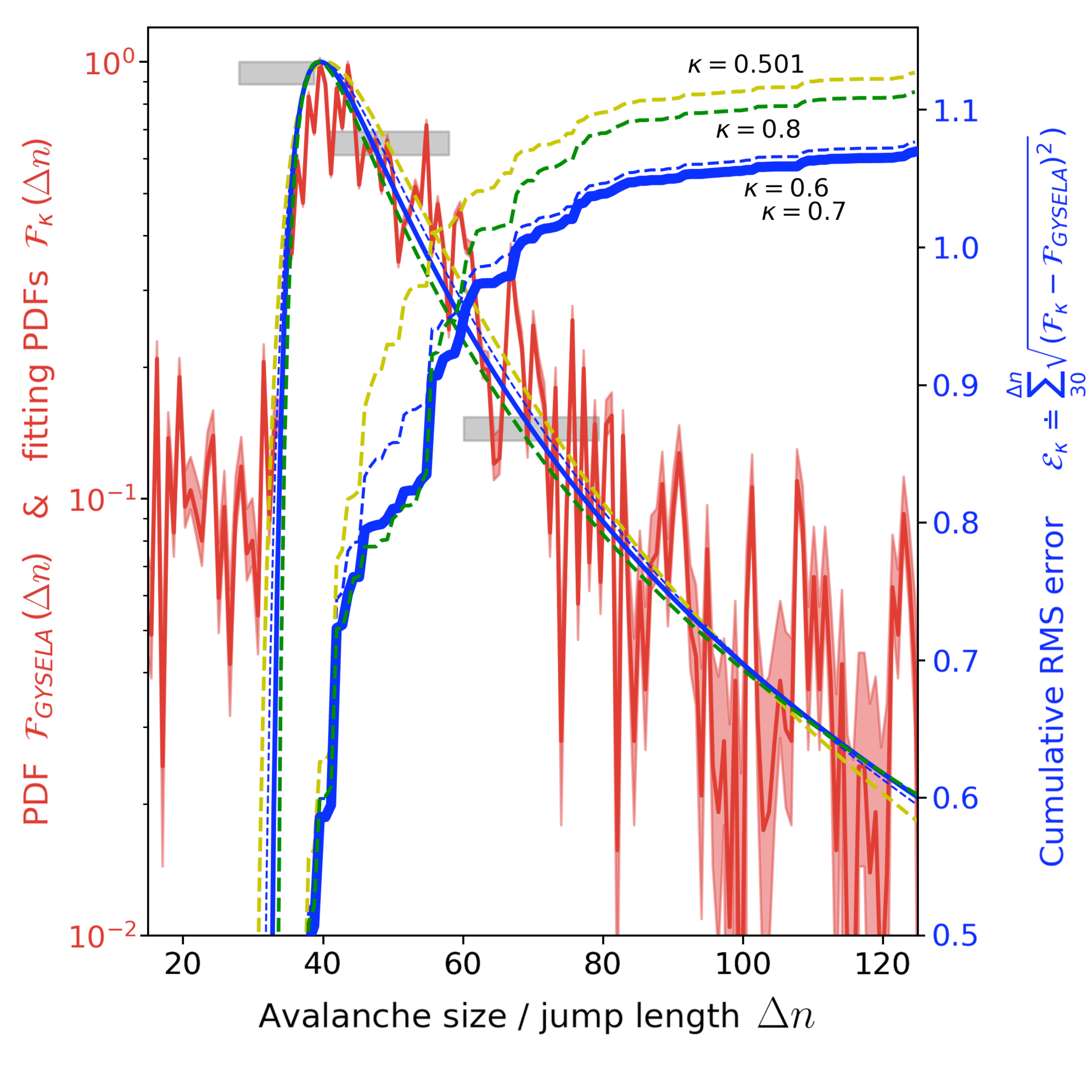}
\caption{\label{} The \textsc{Gysela} computed probability density (the noiselike curve$-$shown in red (light gray) color) against the Fr\'echet distributions (smooth humped curves) for the different values of the $\kappa$ parameter (min $\kappa = 0.501$; max $\kappa = 0.8$). The cumulative root-mean-square error between the numerically computed (\textsc{Gysela}) and template (Fr\'echet) distributions is shown in the same color legend and style, and is also marked by the corresponding $\kappa$ value assumed in Eqs.~(\ref{Fresh-G}) and~(\ref{Fresh-G+}). Note that the error curves are the growing curves starting at $\Delta n \simeq 40$. Coarse experimental distribution (here in a discrete form) is represented in by the three thick, horizontal pieces (gray color), and was obtained by concatenating 15 different occurrences of staircase observations in well-diagnosed Tore Supra shots $\#$ 47 670 and $\#$ 47 923.}
\end{figure}
\begin{figure}[t]
\includegraphics[width=0.49\textwidth]{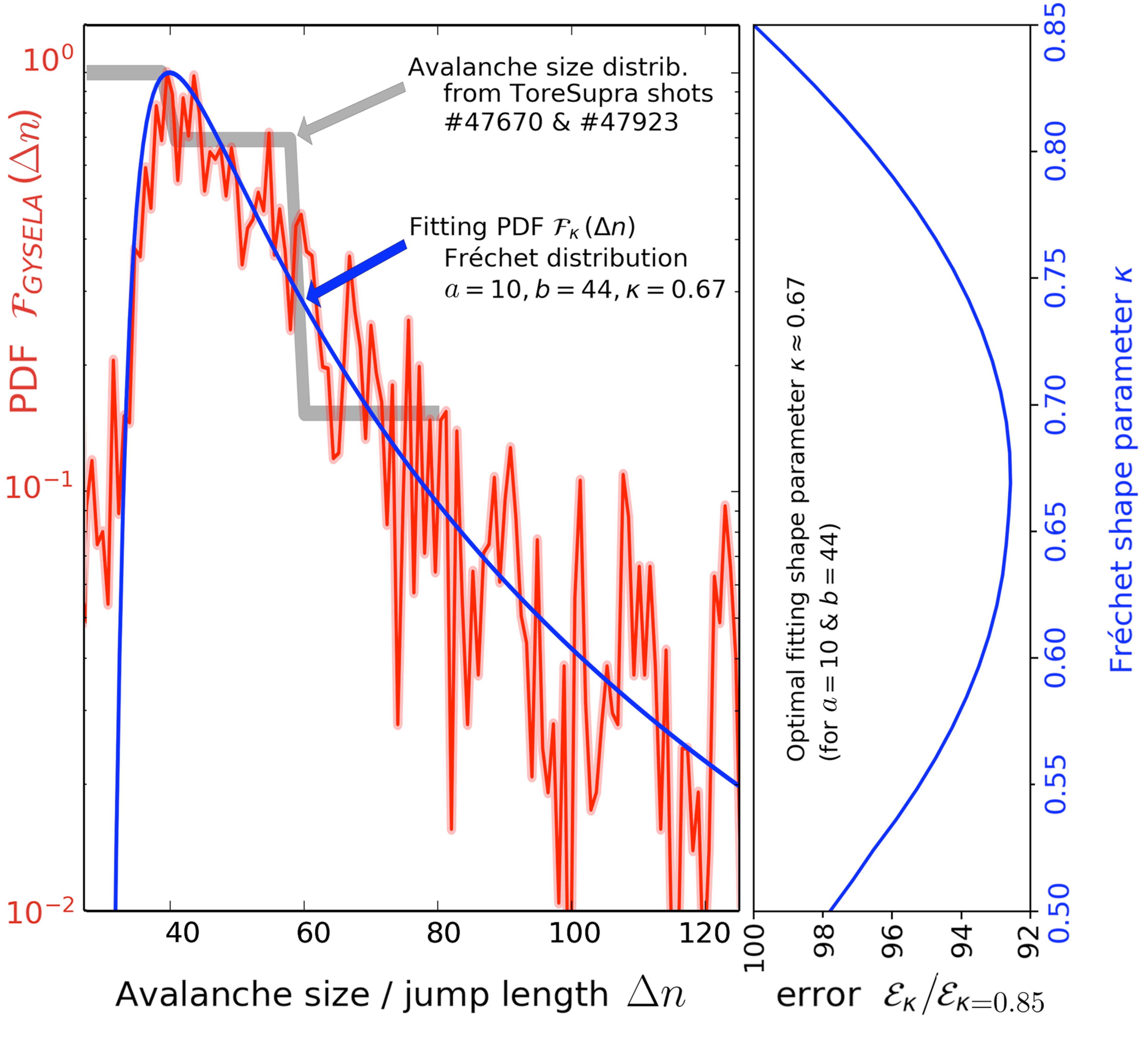}
\caption{\label{} The \textsc{Gysela} computed probability density (the noiselike curve$-$shown in red (light gray) color) versus the Fr\'echet distribution (smooth humped curve) with $\kappa \simeq 0.67$ for which the optimum fit was obtained. The steplike curve (gray color) is coarse experimental distribution and interpolates the discrete distribution as of Fig.~4. The right panel summarizes the normalized root-mean-square error in a percentile to the maximum error at $\kappa \simeq 0.85$, showing that the normalized error is minimized for $\kappa \simeq 0.67$.}
\end{figure}

The fact that the event-size distribution of plasma avalanches proves to follow a power law (see Eq.~(\ref{Fresh_P})) is not really surprising. Indeed the size of an avalanche might be roughly considered as a measure of the energy it carries. According to the NLSE model, the avalanches are generated self-consistently by a noise process, whose origin is attributed to couplings between the nonlinear oscillators in Eq.~(\ref{4s+}). Naturally such oscillators produce a superthermal radiation field acting on cold particles. Then in a basic physics of nonequilibrium processes it is an established result \cite{Mima} that a plasma which is immersed in superthermal radiation undergoes velocity-space diffusion which universally produces an asymptotic power-law distribution over the energies, with a $\kappa$ value which is dictated by the intensity of the radiation (being, as a general rule, inversely proportional to this). In our case this intensity is implicit in the edge condition that the avalanches occur at the localization-delocalization threshold, yielding $\kappa = ({\sqrt{17}} + 1)/8 \simeq 0.64$ in good agreement with the numerical simulation result $\kappa \simeq 0.67$.

\subsection{The case of extreme avalanches}

The fact that the event-size distribution in Eq.~(\ref{Fresh_P}) follows a power law implies that there is an ample space for large-magnitude and extreme avalanches (much ampler at least than what one would expect under a Gaussian hypothesis), and that the extreme avalanches are essentially unpredictable. Indeed the absence of a characteristic scale (other than the finite system size) makes such avalanches indistinguishable from any small- or medium-size avalanche, leading to an impossibility of forecasting \cite{Sornette_2012,Sornette_PRL}. In particular, there might be not any special precursory activity preceding the large-magnitude avalanches, so to a shallow observer it would appear that the large bursts occur so very unexpectedly under seemingly usual plasma conditions. In Fig.~6 we show an example of such large-magnitude event, which was observed in the \textsc{Gysela} simulations following a rather long plasma quiescent period. On the top of the graph we also plotted the flux-surface averaged temperature profiles for the different time lags (i.e., for 450, 680, and 910 time units). That the profiles look like unchanged as time progresses means that the burst in Fig.~6 has developed entirely through the self-organization of the plasma (i.e., via internal redistribution of the free energy already available to the staircase), rather than was a consequence of occasional overdriving of the nonlinear avalanche-jet zonal flow system due to an excessive free-energy injection.

\begin{figure}[t]
\includegraphics[width=0.45\textwidth]{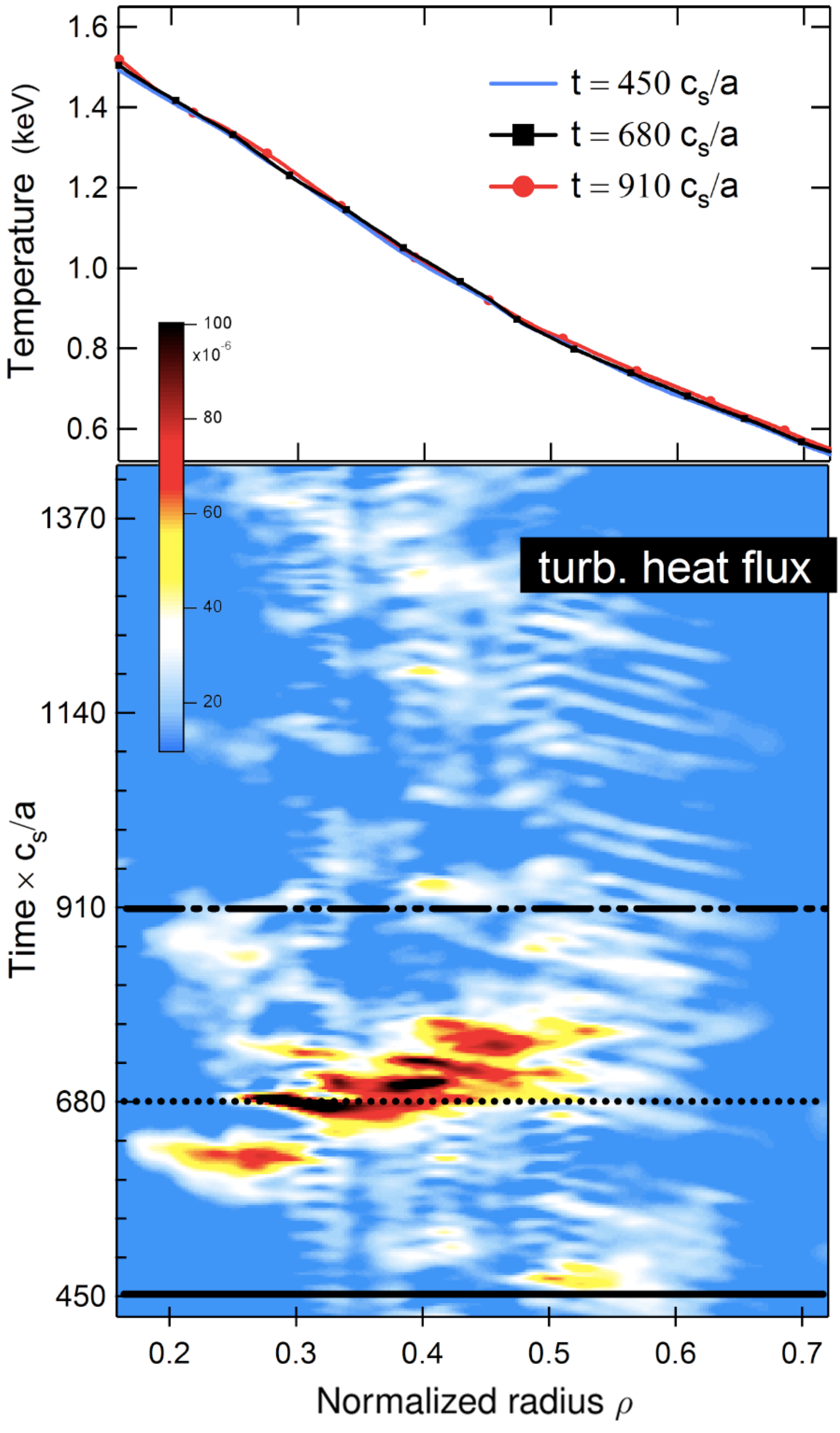}
\caption{\label{} An example of extreme event observed in the \textsc{Gysela} simulations of Tore Supra L-mode plasma. Colors: Blue marks the background heat flux level; white/yellow indicates elevated turbulent flux; dark/brownish signifies extreme flux. Axes: ``$\rm a$" is the tokamak minor radius, ``$\rm c_s$" is the ion acoustic speed, and $\rho$ is the dimensionless radial coordinate. Upper panel: the flux-surface averaged temperature profiles for the different time lags: $t=450, 680$, and 910 time units.%$-$suggesting that the surge had developed entirely through the self-organization of L-mode plasma.
}
\end{figure}

While the occurrence of the large-magnitude and extreme avalanches appears to be a matter of statistics (the fatalist would say is a matter of good luck), a question of practical importance would be to assign a safety grade to a dynamical system with possible extreme events (in our case extreme bursts of transport across the staircase transport barriers). The task is relatively straightforward and refers to an extreme-risk classification scheme already developed in Ref. \cite{CSF} to which the $\tau$ value is key. Comparing the NLSE result $\tau \simeq 2.56$ (our Table II) to the reference values reported in Table 4 of Ref. \cite{CSF}, one might see that the plasma staircase exhibits characteristics enabling to associate it with the safety class $A3$ (moderate tail risk). The class is two grades less than the highest grade achievable for dynamical systems with self-regulation (class $A5$) and three grades less than the safest class (class $X$), which requires external (forced) regulation. While the assignment of the different safety grades might appear to be an academic exercise, we note that the grades are justified from a topological perspective \cite{CSF} and as such might be used to categorize the various dynamical phenomena resulting in extreme events. 

\subsection{Is it SOC or not?}

Based on the above analysis one might reasonably establish that the plasma staircase is a (i) slowly driven; (ii) dissipative; (iii) interaction-dominated; (iv) thresholded nonlinear dynamical system with a (v) large number of interacting degrees of freedom (i.e., coupled nonlinear oscillators in Eq.~(\ref{4s+})). More so, this system is characterized by a power-law event size distribution of dynamical relaxation events (see Eq.~(\ref{Tail+})); by a superdiffusive dispersion relation $t\sim |\Delta n|^\mu$ (which is an immediate consequence of the L\'evy-fractional FFPE~(\ref{FFPE})); and by the criticality character that the stable state minimizes the growth rate of the potential function $\Phi (\Delta n) \propto |\Delta n|^\alpha$, leading to $\alpha = 3$ (Sec.~V\,A). These observations might inspire one to conclude that the plasma staircase is a complex dynamical system in a state of self-organized criticality (SOC). Indeed the requirements (i)$-$(v) have been discussed to be the defining signatures and important key features of SOC (e.g., Refs. \cite{Asch2013,Newman,Asch2016,Sharma}), motivating the above conclusion. The subject had caused some excitation in the literature previously as the theoretical notion of SOC was advanced as a paradigmatic framework to explain the behavior of driven, dissipative dynamical systems in response to slow driving \cite{Works,Jensen,Sornette2004}.  

This said, the exponent $\tau \simeq 2.56$ (safety class $A3$ \cite{CSF}) appears to be numerically very different from the respective values deriving from sand-pile SOC models \cite{Bak87,Tang} and their generalizations (e.g., the Zhang model \cite{Zhang}; the e-pile model \cite{CSF}; and other models alike): If only in one dimension, which is a very special case of SOC \cite{Kadanoff98} (safety class $A1$: high risk of extreme avalanches) as well as in the mean-field limit \cite{Vesp} for which a crossover to the diffusive transport could be expected (safety class $A5$: moderate to low risk of extreme events). 

These differences in safety classes suggest that in the parameter range of staircase self-organization we actually encounter a different type of SOC phenomenon, which is not reproduced by the familiar sand-pile style models. A characteristic feature of this new type is that the avalanches are driven directly by the white L\'evy noise, which is self-organized, and whose origin is found in multiple dynamical couplings between the nonlinear oscillators in Eq.~(\ref{4s+}). At a kinetic level, the avalanches correspond to a L\'evy-fractional FFPE~(\ref{FFPE}) with self-consistent potential and must be considered as coherent structures. Conversely, sand-piles lead to a different description as they rely on next-neighborlike interactions which are local in space and which may generate some non-Markovianity for $t\rightarrow+\infty$, but not really a nonlocal behavior in the sense of the L\'evy-Gnedenko generalized central limit theorem \cite{Gnedenko} (remark that the e-piles \cite{CSF} bring directly to a non-Markovian diffusion equation with memory and the fractional differentiation over {\it time}, while the {\it space} derivatives are integer and correspond to the familiar Laplacian operator, in contrast to FFPE~(\ref{FFPE})). 

Last but not least, we have seen that the marginally stable (probably SOC) state occurs for $s = ({\sqrt{17}} - 1)/{4}$ exactly. We consider this exact value as a mathematical constant characterizing the staircase self-organization. The self-organized nature of the criticality stems from the fact that the cubic dependence in $\Phi (\Delta n) \propto |\Delta n|^3$ attracts the nonlinear avalanche-jet zonal flow dynamics (by minimizing the free energy stored in the staircase). Note that the avalanches undergo weak localization by the potential field $\Phi (\Delta n)$, at no contradiction with the fact that the staircase system can generate extreme avalanches with the significant (far beyond the Gaussian expectation) likelihood.   

\subsection{Subdiffusion of transport barriers}

Let us now revisit the statement at the end of Sec.~III that there is no transport of waves in NLSE~(\ref{1}), if $s < 1$. We repeat ourselves in saying that this forceful statement applies to classical waves for which the discontinuities of the escape path to infinity act as the topological barriers in phase space impeding field-spreading to long distances. In a complex system with a broad fluctuation spectrum, however, it is virtually unavoidable that there is a certain population of low-frequency, long-wavelength modes for which the classical approach as of NLSE~(\ref{1}) would appear to be too crude$-$actually calling for a full operator (i.e., ``quantum") description instead. Of the effects that resist the discontinuity limitations \cite{PRE14} imposed by $s < 1$, we mention under-barrier propagation and other tunneling processes alike, yet omitted the important possibility that nonlinear structures may develop. In the full operator description, the quantum analogue of NLSE~(\ref{1}) predicts that there is a subdiffusive spreading of the nonlinear wave packet in accordance with a sublinear dispersion law (see Eq.~(34) of Ref. \cite{PRE19})      
\begin{equation}
\langle(\Delta n)^2 (t)\rangle \simeq t^{1/(s+1)}. 
\label{Q-Sub} % Eq.~(\ref{Q-Sub})
\end{equation}  
The scaling in Eq.~(\ref{Q-Sub}) is obtained by combining the quantum generalization of dynamical equations~(\ref{eq+s}) with Fermi's golden rule \cite{Golden} for transitions between states. For $s \rightarrow 1$, the behavior on the right-hand side of Eq.~(\ref{Q-Sub}) is square-rootlike leading to the familiar ``half-diffusion," i.e., $\langle(\Delta n)^2 (t)\rangle \simeq t^{1/2}$. This scaling finds support in results of direct numerical simulations of quantum NLSE dynamics based on the Hubbard model \cite{Ivan}. Using for $s$ the SOC value $s = ({\sqrt{17}} - 1)/{4}$, from Eq.~(\ref{Q-Sub}) we have a more precise estimate $1/(s+1) = ({\sqrt{17}} - 3)/{2} \simeq 0.56$. 

In the staircase self-organization, the time dependence in Eq.~(\ref{Q-Sub}) corresponds to a stochastic spreading of the staircase jets along $n$ as a result of their nonlinear interaction. This spreading is actually very slow$-$subdiffusive. By the time this paper is being written, we have no experimental indication whether such a spreading might or might not be the case in L-mode tokamak plasma. {\it A priori} we might expect the spreading law in Eq.~(\ref{Q-Sub}) to apply at relatively long wavelengths not shorter at least than the Rhines length \cite{Naulin} for the electrostatic drift-wave turbulence. This, together with the fact that the electrostatic Rhines length, $\Lambda_{\rm Rh}$, determines the spacing between the jets, might suggest an interesting scenario for the decay of the staircase, according to which the staircase jets would migrate in the direction of the tokamak minor radius until they merge together into singular structures$-$a process favored by the inverse energy cascade in the turbulence domain. If this scenario is true, then one might also predict that the lifetime of the staircase would scale with the Rhines length as $\Delta t_{\rm life} \simeq \Lambda_{\rm Rh}^{2(s+1)} \simeq (E/B)^{s+1}$, implying a rather strong dependence on the radial electric field $E$ (ironically, stronger turbulence implies longer lifetimes). The crucial question is whether the staircases evolve naturally in the direction of this scaling law, or instead their eventual dissipation is governed by processes like Coulomb collisions and/or quasilinear diffusion on microscopic scales.

\section{Summary and Conclusions}

Our work addresses several important questions concerning the physics of plasma staircase. Firstly, we have argued that the plasma staircase operates as a wave packet of coupled nonlinear oscillators, the jet zonal flows, interacting with each other by emitting and absorbing the plasma avalanches and voids. While a universal analytical method that is valid in all aspects of the staircase dynamics would be an impossible task, we could nevertheless formulate a simplified yet relevant theoretical approach based on a modified NLSE with a subquadratic power nonlinearity \cite{PRE19}. Dealing with the subquadratic power has led us to explore mathematical methods that were not quite common with the fusion physicists$-$among these methods were Diophantine equations and the formalism of backbone map \cite{PRE14}. 

Theoretically, the subquadratic nonlinearity proves to be a very appealing type of nonlinear coupling process as it leads directly to the white L\'evy noise in a system with distributed interactions. In a self-consistent description, this noise process acts as an input driving force for radial transport by plasma avalanches. We considered the avalanches as coherent structures driven by complex processes of mode coupling in magnetically confined fusion plasma consistently with the implication of an NLSE. 

Arguing that the particles could be trapped and convected by the avalanches, a probabilistic picture of the microscopic transport has been drawn using a L\'evy-fractional Fokker-Planck equation with self-consistent potential field (we supplemented the self-consistent equation with the familiar Brownian diffusion due to Coulomb collisions and other frictional processes alike; as well as by the apposite sources and sinks to assess the behavior near the origin). This model description is very nontrivial, as it brings the nonlocality, contained in the L\'evy-fractional derivatives, in contact with the nonlinearity, contained in the potential field. 

Mathematically, the nonlocal (Riesz) derivative occurring in FFPE~(\ref{FFPE}) (and the associated white L\'evy noise driving the transport) is a direct consequence of the competing nonlocal ordering assumed in NLSE~(\ref{1}) in terms of the subquadratic power exponent ($s < 1$). In the absence of a competing ordering the resulting transport equation would be obviously local in space, with the Laplacian operator substituting the Riesz operator in Eq.~(\ref{Def+}).  One sees that it is the subquadratic NLSE~(\ref{1}), with $s < 1$, which is the relevant equation to understand the origin of nonlocal transport (through couplings between waves), and not the familiar, quadratic NLSE, with $s=1$.   

Based on the idea that the plasma staircase resides at a state of marginal stability, we could predict the shape of the potential function at the marginality, and eventually obtain the whole set of power exponents characterizing the plasma staircase. In particular, we have found that the plasma staircase generates a power-lawlike event-size distribution of plasma avalanches, with a room for large-magnitude and extreme events, and we have supported this conclusion by results from direct numerical simulations using the \textsc{Gysela} code. 

In general, the simulations have confirmed that the staircase self-organization is inherent to L-mode plasma, being especially very clear when the turbulence is near-critical, i.e., when the turbulence drive is close to or slightly above the linear instability threshold. Also we have seen that the state of marginal stability bears signatures enabling to associate it with a complex system in a state of self-organized criticality, or SOC. This said, the critical exponents, which we obtained, were not consistent with the critical exponents of major sand-pile SOC models and their generalizations, suggesting that the plasma staircase belongs to a different class of SOC. The critical state is characterized by the self-consistently occurring white L\'evy noise driving the plasma avalanches through a grid of self-conistently generated bulk transport barriers (the jets of the staircase), and mathematically corresponds to a very nontrivial value of the subquadratic power exponent, i.e., $s = ({\sqrt{17}} - 1)/{4} \simeq 0.78$. This fancy value is an exact result of the NLSE model. 

In the vicinity of the criticality, the plasma avalanches undergo weak localization, while the asymptotic probability density decays as a power law, with finite second moments. It is understood that the finiteness of the moments is imposed directly by the self-consistent potential field at the edge of the localization-delocalization transition. This finding supports the idea \cite{PRE18,Ch2007} that a L\'evy flyer could be confined within a semi-transparent transport barrier, even though the escape probability appears to be so large that the outward flux goes, still, quite beyond the Gaussian level. 

While the main focus of the present study was on the classical model in NLSE~(\ref{1}), an extension towards a full quantum (operator-based) equation has been addressed. Of the main physics consequences deriving from this extension we mentioned the possibility for the staircase transport barriers to diffuse along the tokamak minor radius in accordance with the subdiffusive scaling law $\langle(\Delta n)^2 (t)\rangle \simeq t^{0.56}$ for $t\rightarrow+\infty$. It is not possible, for the moment being, to prove or disprove this scaling law based on evidence from experiments or from gyrokinetic numerical simulations. Also it is not yet clear if the very phenomenon of subdiffusion of the transport barriers is there for L-mode plasma. Analysis in this general area remains to be carried out.  

All in all, we have seen that modern statistical physics has an important contribution to make in understanding the formation of the plasma staircase, and we expect the analytic methods, devised in this work, to guide further progresses in the study of strongly coupled dynamical systems with nonlocal ordering.

\acknowledgments
AVM is indebted to Alexander Iomin for illuminating discussions on the various aspects of Anderson localization and NLSE dynamics. This work was carried out within the framework of the EUROfusion Consortium and received funding from the Euratom research and training programme 2014-2018 and 2019-2020 under Grant Agreement No. 633053 (Projects No. AWP17-ENR-ENEA-10 and WP19-ER/ENEA-05). Partial support was received from the National Science Foundation under Grant No. NSF PHY-1748958. %In Russia, the work was supported by the academic program ``Plasma" of the Russian Academy of Sciences. 
The work was also granted access to the HPC resources of TGCC and CINES made by GENCI, and to the EUROfusion High Performance Computer Marconi-Fusion. We acknowledge PRACE for awarding us access to Joliot-Curie at GENCI@CEA, France and MareNostrum at Barcelona Supercomputing Center (BSC), Spain. %Partial support was received from the academic program ``Plasma" of the Russian Academy of Sciences. 
%The views and opinions expressed herein do not necessarily reflect those of the European Commission. 

%This work was also granted access to the HPC resources of TGCC and CINES made by GENCI, and to the EUROfusion High Performance Computer Marconi-Fusion. We acknowledge PRACE for awarding us access to Joliot-Curie at GENCI@CEA, France and MareNostrum at Barcelona Supercomputing Center (BSC), Spain. 

\appendix*

\section{Derivation of the L\'evy-fractional Fokker-Planck equation} 

Consider a Markov stochastic process defined by the evolution equation
\begin{equation}
f(n, t+\Delta t) = \int_{-\infty}^{+\infty} f(n-\Delta n, t) \Upsilon (n, \Delta n, \Delta t)d\Delta n,
\label{1+A} % Eq.~(\ref{1+A})
\end{equation}
where $f (n, t)$ is the probability density of finding a particle (random walker) at time $t$ at point $n$, and $\Upsilon (n, \Delta n, \Delta t)$ is the transition probability density of the process. Note that the ``density" $\Upsilon (n, \Delta n, \Delta t)$ is defined with respect to the increment space characterized by the variable $\Delta n$. It may include a parametric dependence on $n$, when non-homogeneous systems are considered. Here, for the sake of simplicity, we restrict ourselves to the homogeneous case, and we omit the $n$ dependence in $\Upsilon (n, \Delta n, \Delta t)$ to enjoy
\begin{equation}
f(n, t+\Delta t) = \int_{-\infty}^{+\infty} f(n-\Delta n, t) \Upsilon (\Delta n, \Delta t)d\Delta n.
\label{2+A} % Eq.~(\ref{2+A})
\end{equation} 
Then $\Upsilon (\Delta n, \Delta t)$ defines the probability density of changing the spatial coordinate $n$ by a value $\Delta n$ within a time interval $\Delta t$ independently of the running $n$ value. The integral on the right of Eq.~(\ref{2+A}) is of the convolution type. In the Fourier space this becomes
\begin{equation}
\widehat f(q, t+\Delta t) = \widehat f(q, t) \widehat \Upsilon (q, \Delta t), 
\label{3+A} % Eq.~(\ref{3+A})
\end{equation} 
where the integral representation  
\begin{equation}
\widehat \Upsilon (q, \Delta t) = \hat \mathrm{T}_q \{\Upsilon (\Delta n, \Delta t)\} \equiv \int_{-\infty}^{+\infty} \Upsilon (\Delta n, \Delta t) e^{iq\Delta n} d\Delta n
\label{Fourier+A} % Eq.~(\ref{Fourier+A})
\end{equation} 
has been used for $\widehat \Upsilon (q, \Delta t)$, and similarly for $\widehat f(q, t)$. In the above $q$ denotes the position coordinate in the Fourier space, $\hat \mathrm{T}_q$ is the operator of the integral transform, and we have introduced a wide hat to mark the resulting Fourier components (a narrow hat is reserved for the operators). Letting $q\rightarrow 0$ in Eq.~(\ref{Fourier+A}), it is found that 
\begin{equation}
\lim_{q\rightarrow 0}\widehat \Upsilon (q, \Delta t) = \int_{-\infty}^{+\infty} \Upsilon (\Delta n, \Delta t) d\Delta n.
\label{F2} % Eq.~(\ref{F2})
\end{equation} 
The improper integral on the right hand side is none other than the probability for the space variable $n$ to acquire {\it any} increment $\Delta n$ during time $\Delta t$. For memoryless stochastic processes without trapping, this probability is immediately seen to be equal to 1, that is, the diffusing particle takes a displacement anyway in any direction along the $n$-axis. Therefore,
\begin{equation}
\lim_{q\rightarrow 0}\widehat \Upsilon (q, \Delta t) = 1.
\label{F2+} % Eq.~(\ref{F2+})
\end{equation} 
We consider $\widehat \Upsilon (q, \Delta t)$ as the average time- and wave-vector dependent transition ``probability" or the characteristic function of the stochastic process in Eq.~(\ref{2+A}). In general, $\widehat \Upsilon (q, \Delta t)$ can be due to many co-existing, independent dynamical processes, each characterized by its own (partial) transition probability, $\widehat \Upsilon_h (q, \Delta t)$, where $h=1,\dots, m$ is an integer counter, making it possible to factorize
\begin{equation}
\widehat \Upsilon (q, \Delta t) = \prod_{h=1}^m \widehat \Upsilon_h (q, \Delta t).
\label{Prod+A} % Eq.~(\ref{Prod+A})
\end{equation} 
We should stress that, by their definition as Fourier integrals, $\widehat \Upsilon_h (q, \Delta t)$ are given by complex functions of the wave vector $q$, and their interpretation as ``probabilities" has the only purpose of factorizing in Eq.~(\ref{Prod+A}). This factorized form is justified via the asymptotic matching procedure in the limit $q\rightarrow 0$. Without losing in generality, it is sufficient to analyze a simplified version of Eq.~(\ref{Prod+A}) with only two processes included$-$one corresponding to a white noiselike process which we shall mark by the index $\mu$; and the other one corresponding to a regular convection process due to the presence of a potential force which we shall mark by the index $R$. We have, accordingly,  
\begin{equation}
\widehat \Upsilon (q, \Delta t) = \widehat \Upsilon_{\mu}  (q, \Delta t) \widehat \Upsilon_{R} (q, \Delta t).
\label{Prod2+A} % Eq.~(\ref{Prod2+A})
\end{equation}
These settings correspond to a set of Langevin equations
\begin{equation}
dn/dt = v;~dv/dt = -\eta v + F_R + F_\mu (t),
\label{Lvin+A} % Eq.~(\ref{Lvin+A})
\end{equation}
where $\eta$ is the fluid viscosity; $v$ is the velocity along the $n$ axis; $F_R$ is the regular force; and $F_\mu (t)$ is the fluctuating (noiselike) force. We take $F_\mu (t)$ to be a white L\'evy noise with L\'evy index $\mu$ ($1 < \mu\leq 2$). By white L\'evy noise $F_\mu (t)$ we mean a stationary random process, such that the corresponding motion process, i.e., the time integral of the noise, $L_\mu (\Delta t) = \int_t^{t+\Delta t} F_\mu (t^{\prime}) dt^{\prime}$, is a symmetric $\mu$-stable L\'evy process with stationary independent increments and the characteristic function \cite{Report,Klafter} 
\begin{equation}
\widehat \Upsilon_\mu (q, \Delta t) = \exp (-D_\mu |q|^\mu \Delta t) \simeq 1 - D_\mu |q|^\mu \Delta t.
\label{GCLT+A} % Eq.~(\ref{GCLT+A})
\end{equation}
The last term gives an asymptotic inverse-power distribution of jump lengths 
\begin{equation}
\chi (\Delta n) \sim |\Delta n|^{-(1+\mu)}.
\label{Jump-l+A} % Eq.~(\ref{Jump-l+A})
\end{equation}
In the above, the coefficient $D_\mu$ is the intensity of the L\'evy noise. As is well-known, the characteristic function in Eq.~(\ref{GCLT+A}) generates L\'evy flights \cite{Ch2007,Klafter,Georges}. 

Focusing on the regular component of the force field, $F_R$, it is convenient to represent the corresponding transition probability in the form of a plane wave, i.e.,
\begin{equation}
\widehat \Upsilon_R (q, \Delta t) = \exp (iqu\Delta t) \simeq 1 + iqu\Delta t.
\label{Plane+A} % Eq.~(\ref{Plane+A})
\end{equation}
Here, $u$ is the speed of the wave, and $qu$ is the frequency. As usual, one evaluates the speed $u$ by balancing the regular force $F_R$ to the viscous term in the Langevin Eq.~(\ref{Lvin+A}), yielding $u = F_R/\eta$. It is noted that the basic condition in Eq.~(\ref{F2+}) is well satisfied for both the L\'evy processes and stationary convection, just highlighting the Markov property and the absence of trapping. Putting all the various pieces together, one obtains 
\begin{equation}
\widehat \Upsilon (q, \Delta t) = \exp (-D_\mu |q|^\mu \Delta t + iq F_R \Delta t / \eta),
\label{Tog+A} % Eq.~(\ref{Tog+A})
\end{equation} 
from which Eq.~(\ref{F2+}) is evident. The next step is to substitute Eq.~(\ref{Tog+A}) into~(\ref{3+A}), and to allow $\Delta t \rightarrow 0$. Then, Taylor expanding on the left- and right-hand sides in powers of $\Delta t$, and keeping first non-vanishing orders, in the long-wavelength limit $q\rightarrow 0$ it is found that 
\begin{equation}
\frac{\partial}{\partial t} \widehat f (q, t) = \left[-D_\mu |q|^\mu + iq F_R / \eta \right] \widehat f (q, t).
\label{9+A} % Eq.~(\ref{9+A})
\end{equation} 
When inverted to configuration space, the latter equation becomes
\begin{equation}
\frac{\partial}{\partial t} f (n, t) =  \left[D_\mu \frac{\partial^\mu}{\partial |n|^\mu} - \frac{1}{\eta} \frac{\partial}{\partial n} F_R \right] f (n, t),
\label{Inv+A} % Eq.~(\ref{Inv+A})
\end{equation} 
where the symbol $\partial^{\mu} / \partial |n|^\mu$ is defined by its Fourier transform as  
\begin{equation}
\hat \mathrm{T}_q \Big\{\frac{\partial^\mu}{\partial |n|^\mu}f (n, t)\Big\} = -|q|^\mu \widehat f (q,t).
\label{Def+A} % Eq.~(\ref{Def+A})
\end{equation} 
In the foundations of fractional calculus (e.g., Refs. \cite{Podlubny,Samko}) it is shown that, for $1 <\mu < 2$, 
\begin{equation}
\frac{\partial^\mu}{\partial |n|^\mu} f (n, t) = \frac{1}{\Gamma_\mu}\frac{\partial^2}{\partial n^2} \int_{-\infty}^{+\infty}\frac{f (n^\prime, t)}{|n-n^\prime|^{\mu - 1}} dn^\prime.
\label{Def+A} % Eq.~(\ref{Def+A})
\end{equation} 
Equation~(\ref{Def+A}) reproduces the Riesz fractional derivative in Eq.~(\ref{Def+}), with $\Gamma_\mu = - 2\cos(\pi\mu/2)\Gamma(2-\mu)$. 

Relating $F_R$ to external potential field with the aid of $F_R = -\Phi^{\prime}(n)$, and substituting in Eq.~(\ref{Inv+A}), one arrives at the following L\'evy-fractional Fokker-Planck equation in the $n$-space
\begin{equation}
\frac{\partial}{\partial t} f (n, t) =  \left[D_\mu \frac{\partial^\mu}{\partial |n|^\mu} + \frac{1}{\eta} \frac{\partial}{\partial n} \Phi^{\prime}(n)\right] f (n, t),
\label{FFPE+A} % Eq.~(\ref{FFPE+A})
\end{equation} 
which results in Eq.~(\ref{FFPE}) of Sec. III. 

If we used, in place of Eq.~(\ref{GCLT+A}), the characteristic function of the Brownian white noise, i.e., 
\begin{equation}
\widehat \Upsilon_B (q, \Delta t) = \exp (-D q^2 \Delta t) \simeq 1 - D q^2 \Delta t,
\label{GCLT+AA} % Eq.~(\ref{GCLT+AA})
\end{equation}
we would have obtained, instead of FFPE~(\ref{FFPE+A}), the familiar Fokker-Planck equation 
\begin{equation}
\frac{\partial}{\partial t} f (n, t) =  \left[D \frac{\partial^2}{\partial n^2} + \frac{1}{\eta} \frac{\partial}{\partial n} \Phi^{\prime}(n)\right] f (n, t),
\label{FFPE+AA} % Eq.~(\ref{FFPE+AA})
\end{equation} 
where the stochastic spreading of the probability density corresponds to the second-order derivative over the coordinate $n$, and the coefficient $D$ is the intensity of the Brownian noise. 

Another situation of interest here is when the L\'evy and Brownian noises are present jointly on an equal footing. In that case Eq.~(\ref{Prod+A}) becomes a product of three terms, i.e., the L\'evy term, the Brownian term, and the convection term, yielding, instead of Eq.~(\ref{Prod2+A}),
\begin{equation}
\widehat \Upsilon (q, \Delta t) = \widehat \Upsilon_{\mu}  (q, \Delta t) \widehat \Upsilon_{B}  (q, \Delta t) \widehat \Upsilon_{R} (q, \Delta t).
\label{Prod2+AA} % Eq.~(\ref{Prod2+AA})
\end{equation}
The Langevin equations in Eq.~(\ref{Lvin+A}) generalize to
\begin{equation}
dn/dt = v;~dv/dt = -\eta v + F_R + F_\mu (t) + F_B (t),
\label{Lvin+AA} % Eq.~(\ref{Lvin+AA})
\end{equation}
where $F_B (t)$ denotes the Brownian noise and is added to the L\'evy noise. Substituting the known characteristic functions for the L\'evy and Brownian noises into Eq.~(\ref{Prod2+AA}), and going through steps of the derivation, one arrives at the hybrid FFPE~(\ref{FFPE+}) with both the fractional (Riesz) and ordinary diffusion terms, weighted by the coefficients $D_\mu$ and $D$, i.e.,  
\begin{equation}
\frac{\partial}{\partial t} f (n, t) =  \left[D_\mu \frac{\partial^\mu}{\partial |n|^\mu} + D \frac{\partial^2}{\partial n^2} + \frac{1}{\eta} \frac{\partial}{\partial n} \Phi^{\prime}(n)\right] f (n, t).
\label{FFPE+AA} % Eq.~(\ref{FFPE+AA})
\end{equation} 
Note that FFPE~(\ref{FFPE+AA}) involves space fractional differentiation only in terms of the generalized diffusion operator; whereas the convection term due to $\Phi (n)$ is {\it integer} and introduces a potential well for L\'evy flights. This last observation elucidates the fundamentally different roles played by respectively the stochastic and regular forces as they join together to set up the analytical structure of FFPE.

%
%
%
%


\begin{thebibliography}{}
\bibitem{Comm-Res}
Conditions for nonlinear resonance are obtained by accounting for the nonlinear frequency shift.
\end{thebibliography}

\end{document}